\documentclass[fleqn,usenatbib]{mnras}

\usepackage[T1]{fontenc}

\DeclareRobustCommand{\VAN}[3]{#2}
\let\VANthebibliography\thebibliography
\def\thebibliography{\DeclareRobustCommand{\VAN}[3]{##3}\VANthebibliography}

\usepackage{graphicx}
\usepackage{dcolumn}
\usepackage{bm}
\usepackage{amsmath}
\usepackage{amssymb}
\usepackage{appendix}
\usepackage{color}
\usepackage{soul}
\usepackage{mathtools}
\usepackage{changes}
\usepackage{amsfonts}
\usepackage{tikz}
\usepackage{relsize}
\usepackage{rotating}
\usepackage{xcolor}
\usepackage{widetext}
\usepackage{pdflscape}
\usepackage{threeparttable}
\definecolor{darkgreen}{rgb}{0.0,0.5,0.0}

\input epsf

\usepackage[makeroom]{cancel}
\DeclareMathOperator*{\MyEqual}{=}
\DeclareMathOperator*{\MyEquiv}{\equiv}

\newcommand{\ie}{\emph{i.e.} }
\newcommand{\eg}{\emph{e.g.,} }

\newcommand{\be}{\begin{equation}}
\newcommand{\ee}{\end{equation}}
\newcommand{\bea}{\begin{equation*}}
\newcommand{\eea}{\end{equation*}}
\newcommand{\beqr}{\begin{eqnarray} \nonumber}
\newcommand{\eeqr}{\end{eqnarray}}
\newcommand{\beqrb}{\begin{eqnarray}}
\newcommand{\eeqrb}{\nonumber \end{eqnarray}}
\newcommand{\fin}{\mbox{ .}}
\newcommand{\coma}{\mbox{ ,}}

\newcommand{\yr}{\mbox{ yr}}

\newcommand{\MHz}{\mbox{ MHz}}
\newcommand{\GHz}{\mbox{ GHz}}

\newcommand{\km}{\mbox{ km}}
\newcommand{\pc}{\mbox{ pc}}

\newcommand{\TeV}{\mbox{ TeV}}

\newcommand{\muG}{\mbox{ $\mu$G}}

\newcommand{\const}{\mbox{const}}

\newcommand{\gama}{$\gamma$}

\newcommand{\energain}{g}

\newcommand{\myPret}{P_{\tiny\text{ret}}}
\newcommand{\mysiso}{s_{\tiny\text{iso}}}
\newcommand{\myPesc}{P_{\tiny\text{esc}}}
\newcommand{\RPWN}{R_{\mbox{\tiny PWN}}}
\newcommand{\tdip}{t_{\mbox{\tiny dip}}}
\newcommand{\qdinf}{q_{\infty}}
\newcommand{\kyr}{{\mbox{ kyr}}}

\newcommand{\TS}{\mathcal{TS}}
\newcommand{\termshock}{TS}
\newcommand{\isoc}{\mbox{\tiny ISO}}
\newcommand{\difcases}[1]{\mbox{\tiny {#1}}}

\makeatletter
\DeclareFontFamily{OMX}{MnSymbolE}{}
\DeclareSymbolFont{MnLargeSymbols}{OMX}{MnSymbolE}{m}{n}
\SetSymbolFont{MnLargeSymbols}{bold}{OMX}{MnSymbolE}{b}{n}
\DeclareFontShape{OMX}{MnSymbolE}{m}{n}{
    <-6>  MnSymbolE5
   <6-7>  MnSymbolE6
   <7-8>  MnSymbolE7
   <8-9>  MnSymbolE8
   <9-10> MnSymbolE9
  <10-12> MnSymbolE10
  <12->   MnSymbolE12
}{}
\DeclareFontShape{OMX}{MnSymbolE}{b}{n}{
    <-6>  MnSymbolE-Bold5
   <6-7>  MnSymbolE-Bold6
   <7-8>  MnSymbolE-Bold7
   <8-9>  MnSymbolE-Bold8
   <9-10> MnSymbolE-Bold9
  <10-12> MnSymbolE-Bold10
  <12->   MnSymbolE-Bold12
}{}

\let\llangle\@undefined
\let\rrangle\@undefined
\DeclareMathDelimiter{\llangle}{\mathopen}%
                     {MnLargeSymbols}{'164}{MnLargeSymbols}{'164}
\DeclareMathDelimiter{\rrangle}{\mathclose}%
                     {MnLargeSymbols}{'171}{MnLargeSymbols}{'171}
\makeatother

\title[PWNe spectra and DSA]{Maximally-hard spectra from diffusive shock-acceleration in pulsar-wind nebulae
}

\author[Arad et al.]{
Ofir Arad,$^{1}$\thanks{E-mail: ofirara@post.bgu.ac.il}
Assaf Lavi,$^{1}$
and Uri Keshet$^{1}$\thanks{E-mail: ukeshet@bgu.ac.il}
\\
$^{1}$Physics Department, Ben-Gurion University of the Negev, POB 653, Be'er-Sheva 84105, Israel
}

\date{Accepted XXX. Received YYY; in original form ZZZ}

\pubyear{2020}

\begin{document}
\label{firstpage}
\pagerange{\pageref{firstpage}--\pageref{lastpage}}
\maketitle

\begin{abstract}
The processes leading to the exceptionally hard radio spectra of pulsar-wind nebulae (PWNe) are not yet understood.
Radio photon spectral indices among $29$ PWNe from the literature show an approximately normal, $\alpha=0.2\pm0.2$ distribution.
We present $\sim 3\sigma$ evidence for a distinct sub-population of PWNe, with a hard spectrum $\alpha=0.01\pm0.06$ near the termination shock and significantly softer elsewhere, possibly due to a recent evacuation of the shock surroundings.
Such spectra, especially in the hard sub-population, suggest a Fermi process, such as diffusive shock acceleration, at its extreme, $\alpha=0$ limit.
We show that this limit is approached for sufficiently anisotropic small-angle scattering,
enhanced on either side of the shock for particles approaching the shock front.
In the upstream, the spectral hardening is mostly associated with an enhanced energy gain, possibly driven by the same beamed particles crossing the shock.
Downstream, the main effect is a diminished escape probability,
but this lowers the acceleration efficiency to $\lesssim25\%$ for $\alpha=0.3$ and $\lesssim1\%$ for $\alpha=0.03$.
\end{abstract}

\begin{keywords}
shock waves --- acceleration of particles --- pulsars: general --- magnetic fields
\end{keywords}



\date{\today}
\maketitle

\section{Introduction}
\label{sec:Introduction}

Radio observations of pulsar wind nebulae (PWNe) indicate electron and positron (henceforth electron) distributions with exceptionally hard spectra.
While these electrons appear to be accelerated in the PWN termination shock \citep[\termshock{}; \eg][]{BietenholzEtAl01_Wisps, lyubarsky2003termination}, the spectrum is generally considered too hard for diffusive shock acceleration (DSA).
We analyse the distribution of radio spectra among the $\sim30$ PWNe with available radio data, and study the viability of DSA as the mechanism responsible for energetic electrons in these systems.

\subsection{PWNe structure and evolution}
\label{sec:intro_pwn}

Rotating, highly magnetised neutron stars (NSs) that appear as pulsars are thought to form by the collapse of a massive star associated with a supernova explosion.
The observed slowing-down of the pulsar spin, \ie its increasing spin period $P$ (measured in seconds, henceforth), is known as the spin-down, $\dot{P}>0$.
The inferred, so-called spin-down luminosity, is thought to energise the region surrounding the pulsar \citep{weiler1978crab}, known as the PWN.
We define $\RPWN$ as the typical radius of this nebula, containing most of the synchrotron emission associated with the NS.

The spin-down indicates that the pulsar is losing kinetic energy, providing an estimate of the equatorial dipole field,
\begin{equation} \label{eq:Bzero}
B_{\mbox{\tiny dip}}\simeq  10^{19.5}(P\dot{P})^{1/2}\mbox{ G} \simeq 10^{11}\left(\frac{P}{\mbox{ms}}\right) \left(\frac{\tdip}\kyr\right)^{-1/2}\mbox{G}\coma
\end{equation}
and of the pulsar age, $t<\tdip\equiv P/(2\dot{P})$.
In the aligned-rotator model, where the magnetic dipole
and the NS rotation axis
coincide, the nebular magnetic field at radius $r$ can be crudely related to $B_{\mbox{\tiny dip}}$, \citep{goldreich1969pulsar}
\begin{eqnarray}
    B & \sim &  2\pi^2\frac{R_p^3B_{\mbox{\tiny dip}}}{c^2P^2r} \\
    & \sim &
    \left(\frac{R_p}{10\km}\right)^3
    \left(\frac{\tdip}{\mbox{kyr}}\right)^{-\frac{1}{2}}
    \left(\frac{P}{\mbox{ms}}\right)^{-1} \left(\frac{r}{1\pc}\right)^{-1}
    \mbox{ mG} \coma \nonumber
\end{eqnarray}
where $R_p$ is the NS radius.

In the standard picture, the nebula initially inflates into the outer, slowly-moving supernova remnant (SNR) ejecta.
After a few$\kyr$, the SNR reverse shock collides with the PWN and crushes it \citep{reynolds1984evolution}.
Later, at time $\gtrsim 10^4\yr$, the nebula typically migrates to the SNR periphery, where its motion through the cooling ejecta becomes supersonic.
The resulting bow shock further confines the PWN, and persists after the PWN is ejected from the SNR into the interstellar medium \citep[\eg][]{gaensler2006evolution}.

While the energy released near the NS magnetosphere is dominated by Poynting flux, most of the energy is converted at some distance beyond the light cylinder into a relativistic cold wind \citep{reesgunn1974,kennel1984confinement}. This relativistic wind is slowed down in a \termshock{},
beyond which the hot plasma is radio-bright with synchrotron radiation.
This \termshock{}, of Lorentz factor $\gamma\gtrsim10^4$, is thought to accelerate charged particles to ultra-relativistic energies \citep{gaensler2006evolution,buhler2014surprising}.
The radius $R_w$ of this \termshock{} can be determined by comparing the ram pressure of the wind to the internal pressure of the shocked synchrotron nebula.

\subsection{Synchrotron emission from PWNe}

The observed integrated flux from a PWN constitutes a considerable fraction of the spin-down luminosity of the NS.
In the prototypical example of the Crab nebula, about $\sim25\%$ of the energy dissipated from the pulsar (most of it thought to be in the form of a relativistic wind) is accounted for as radiation emitted from the nebula \citep{hester2008crab}. This emission is widely interpreted as synchrotron radiation from electrons
accelerated in the \termshock{} and gyrating in the PWN magnetic field \citep{shklovskii1953nature,dombrovsky1954nature}.

The broadband emission from PWNe, as in the Crab nebule, ranges from a few $100\MHz$ up to $\sim$TeV energies \citep[][and references therein]{buhler2014surprising}.
For typical $B\sim100\muG$ fields, the synchrotron emission extends from the radio up to the X-ray band,
above which the radiation is attributed to inverse-Compton (IC) emission \citep{grindlay1971compton}.

Denote the specific brightness at frequency $\nu$ as a local power-law of index $\alpha$, $S_{\nu}\propto \nu^{-\alpha}$.
The observed photon spectrum of a typical PWN is characterised by a hard radio spectral index in the range $0\lesssim\alpha_R\lesssim0.3$ (see Fig.~\ref{fig:histogram}), and a softer X-ray spectrum with index $\alpha_X\gtrsim1$.
The spectral softening between these bands,  $\Delta\equiv \alpha_X-\alpha_R\gtrsim0.7$, is somewhat stronger than the theoretical synchrotron cooling-break ($\Delta\simeq0.5$), and
is still not well understood.

The synchrotron cooling time can be estimated as
\begin{eqnarray}
    \label{eq:sych_cool}
    t_{\mbox{\tiny cool}} & \simeq & \frac{3}{\sigma_ T} \sqrt{\frac{2 \pi em_ec}{B^3 \nu }} \\
    & \simeq & 1.3\left(\frac{B}{100\muG}\right)^{-\frac{3}{2}}\left(\frac{\nu}{1~\mbox{GHz}}\right)^{-\frac{1}{2}}~\mbox{Myr}\coma \nonumber
\end{eqnarray}
so one expects a cooling break at a frequency \citep{ginzburg1965cosmic}
\begin{equation}
    \label{eq:spectral_break_frequency}
        \nu_{b}\simeq \frac{18\pi em_ec}{\sigma^2_TB^3t^2}\sim2\times10^{15}\left(\frac{B}{100\muG}\right)^{-3}\left(\frac{t}{\mbox{kyr}}\right)^{-2}\mbox{ Hz}\fin
\end{equation}
Here, $c$ is the speed of light, $\sigma_T$ is the Thomson cross-section, and $m_e$ and $e$ are the electron mass and charge. The radio spectrum of PWNe is therefore largely unaffected by synchrotron cooling.

For instance, in the Crab nebula, for $t\sim1~\mbox{kyr}$, $R_w\sim0.13\pc$, $P\simeq33.6~\mbox{ms}$, and $\dot{P}\simeq4.2\times10^{-13}$ \citep{weisskopf2000discovery,Hester_2002,Abdo_2013} one finds $\tdip\sim 1.3~\mbox{kyr}$ and $B\sim 250\muG$ at $r\simeq R_w$ in the synchrotron nebula, giving $\nu_b\sim 10^{15}~\mbox{Hz}$, consistent with the observed break.
For such a magnetic field, $t_{\mbox{\tiny cool}}\sim300\kyr$ at $\nu=1~\mbox{GHz}$.

We focus on the radio band, where the spectrum is measured over about two decades in photon frequency, and even four decades in the Crab and 3C58 nebulae \citep{buhler2014surprising,Kothes2017},
henceforth dropping the subscript $R$ from $\alpha$.
We define the electron spectral index as $s\equiv-d\ln f/d\ln p$, where $f$ is the particle distribution function (PDF) and $p$ is the particle momentum.
The index $s$ of synchrotron-emitting electrons, well-below the cooling break, is then inferred from the photon spectral index $\alpha$ through the relation $\alpha=(s-3)/2$.
The measured range of $\alpha$ thus indicates spectral indices $3.0\lesssim s\lesssim3.6$ for the radio-emitting electrons, considerably harder than the $s>4$ spectra typically found in shock systems.
Such hard spectra are extreme in the sense that a minute fraction of particles carry most of the energy;
the physical mechanism giving rise to such extreme spectra is not yet understood.

\subsection{Particle acceleration}

Nonthermal, power-law energy distributions of relativistic, high-energy, charged particles, are ubiquitous in astronomy, and are usually associated with acceleration in a collisionless shock.
In DSA, repeated collisions of charged particles with magnetic irregularities scatter the particles back and forth across the shock, indirectly drawing energy from the bulk flow in a first-order Fermi process \citep[but see \eg][for a discussion of alternative acceleration processes in shocks]{AronsTavani94}.

As charged particles around the shock scatter off the magnetic irregularities, they modify the scattering modes themselves, and --- when carrying sufficient energy --- the structure of the shock as well, rendering the problem highly nonlinear.
No self-consistent theory fully accounts for the shock structure, the surrounding magnetic irregularities, and the associated particle acceleration.
An approximate approach is to adopt some ansatz for the form of the scattering function $\kappa$, and examine the resulting particle spectrum, in the so-called test-particle approximation\footnote{The term is somewhat of a misnomer here, as the form of the scattering function is not known in a collisionless shock even when the fraction of energy deposited in relativistic particles is small.}.
In the limit of  a non-relativistic, strong shock, propagating into a medium with an adiabatic index $\Gamma=5/3$, one then finds a flat energy spectrum, $s\simeq 4$, in general agreement with observations, provided that scattering is sufficiently isotropic \citep{keshet2020diffusive}, although non-linear effects could modify the picture \citep[for reviews, see][]{Blandford_Eichler_87, malkov2001nonlinear, CaprioliHaggerty19}.

For relativistic shocks, such as in the PWNe studied here, DSA becomes more complicated, mainly due to the anisotropy of the accelerated particle distribution \citep[for recent reviews, see][]{sironi2015relativistic, pelletier2017}.
The problem has been solved under different assumptions on $\kappa$, in various methods: numerically, using eigenfunction decompositions \citep{kirk1987acceleration, heavens1988relativistic, Kirk_2000}, Monte Carlo simulations, \citep{Bednarz_ostrowski1996, bednarz2000acceleration, achterberg2001particle} and relaxation \citep{nagar2019diffusive} codes, semi-analytically, using an angular moment expansion \citep{Keshet06}, and in an analytic approximation \citep{Keshet_2005}.

In the limit of isotropic, small-angle scattering, one finds a spectral index $s\simeq4.22$ in highly relativistic shocks \citep{heavens1988relativistic, Kirk_2000}, asymptoting to $s=38/9$ in the ultra-relativistic shock limit \citep{Keshet_2005}.
This value is in agreement with \gama-ray burst (GRB) afterglow observations \citep{freedman2001energy, curran2010electron, Fong2015}; see \citet{lavi2020diffusive} for a recent discussion, and with the spectrum of $\gtrsim \TeV$ electrons in PWNe, as inferred from X-ray emission \citep{hillas1998spectrum,galindo2014discovery,aleksic2015measurement}.
The limiting spectrum is robust for isotropic scattering, retaining similar values for some anisotropic choices of $\kappa$  \citep[\eg][]{Kirk_2000} and in other dimensions \citep{keshet2017analytic,lavi2020diffusive}, although non-linear effects could modify the picture if $\kappa$ is not a local function of the PDF \citep{nagar2019diffusive}.
However, the limiting spectrum was found to be quite sensitive to the form of the small-angle scattering function \citep{Keshet06} in dimensions above one \citep{keshet2017analytic}, and can become very hard for some choices of large-angle scattering \citep{ellison1990first,meli2003particle,summerlin2011diffusive}.

A more direct approach to the study of particle acceleration involves ab-initio, kinetic plasma simulations.
Particle in cell (PIC) simulations of shocks have shown the onset of power-law spectrum \citep{spitkovsky2008particle, martins2009radiation,lemoine2010onelectro, sironi2011acceleration,plotnikov2013particle,plotnikov2018perpendicular,lemoine2019,marcowith2020multi}.
However, due to computational limits, such simulations only probe the early stages of particle acceleration, and infer the spectrum of the developing high energy tail from $\lesssim$ two decades in energy, typically in two spatial dimensions.

It is generally thought that
DSA in an ultra-relativistic shock does not naturally generate a spectrum as hard as that observed in PWNe \citep[\eg][]{Kirk_2000,ostrowski2002comment,summerlin2011diffusive}.
Furthermore, DSA in a magnetised relativistic shock is easily quenched as strong perpendicular magnetic fields confine the particles \citep{ballard1991first, kirk1989particle}, as demonstrated using PIC simulations for strong \citep{sironi2009particle} and intermediate-strength \citep{sironi2013maximum} fields.
The efficiency of particle acceleration is thus sensitive to the unknown dissipation of the approximately toroidal, striped field \citep{lyubarsky_2001, kirk2003dissipation, sironi2011acceleration}.

The unusual radio spectrum of PWNe has led to the consideration of alternative acceleration mechanisms, such as the absorption of cyclotron waves \citep{hoshino1992relativistic,GallantArons94} in the presence of sufficient ions \citep{GallantEtAl02, amato2006heating}.
Magnetic reconnection in the striped-wind was explored as a possible acceleration mechanism \citep{kirk2003dissipation, lyubarsky2003termination,petri2012high}, including with PIC simulations \citep[\eg][]{sironi2011acceleration,sironi2012particle,cerutti2014three,cerutti2020global}.
Reconnection in 2D simulations was shown to generate a hard population of electrons, where the spectral index approaches $s\sim3.2$ for highly magnetised shocks \citep{Sironi_2014,werner2015extent,kagan2018}.
Attributing the radio emission to reconnection is challenged, in particular by the narrow, factor $\sim40$ range of energies where the hard spectrum is reproduced \citep{kagan2018}. In addition, 2D simulations predict a narrow thermal-like distribution \citep{sironi2017particle}. For reviews of reconnection in PWNe, see \citet{sironi2015relativistic, kagan2015relativistic,sironi2017particle}.

\subsection{Outline}

In this paper, we analyse radio PWNe from the literature, study the distribution of the implied synchrotron-emitting electrons in different types of nebulae, and examine whether DSA can account for these spectra in the anisotropic, small-angle scattering limit.

The paper is organised as follows. In \S\ref{sec:pwne}, we conduct a statistical analysis of PWN radio spectral indices from the literature, exploring in particular a possible association between PWN spectrum and morphology. 
In \S\ref{sec:method}, we outline the DSA framework, present the moment-based method used to derive the spectrum, and study the energy gain and return probability diagnostics used to characterise the acceleration process.
We explore different models for anisotropic angular diffusion in \S\ref{sec:hard_spectrum}, and examine their effect on the resulting spectrum and PDF.
Finally, we discuss the possible adequacy of DSA scenarios to the hard spectra observed in PWNe.
The results are summarised and discussed in \S\ref{sec:summary}.
Details pertaining to the specific DSA models are provided in \S\ref{sec:appendix_cases}.

\section{Distribution of PWN spectra}
\label{sec:pwne}

In this section we introduce and review all available PWNe spectral indices from the literature.
After introducing the data and analysis method (in \S\ref{sec:data_prep}), as summarised in Table \ref{pwn_table}, we model the distribution of spectral indices (\S\ref{subsec:Model}), and consider (\S\ref{sec:classifications}) the possible presence of a distinctively hard sub-population of PWN associated with exposed cores, which has interesting possible implications (\S\ref{subsec:Implications}).

\subsection{Radio spectra of literature PWNe}
\label{sec:data_prep}

We analyse all 31 PWNe with a measured radio spectrum we could find reported in the literature up to 2019, as summarised in Table~\ref{pwn_table}.
The spectra are presented both as photon spectral indices $\alpha$ (along with the reported uncertainty, when available) and as the corresponding electron spectral indices $s=3+2\alpha$.
The table provides basic information for each PWN, including the age, morphological classification, and associated pulsar when available.

\begin{figure}
\centerline{\epsfxsize=8.4cm \epsfbox{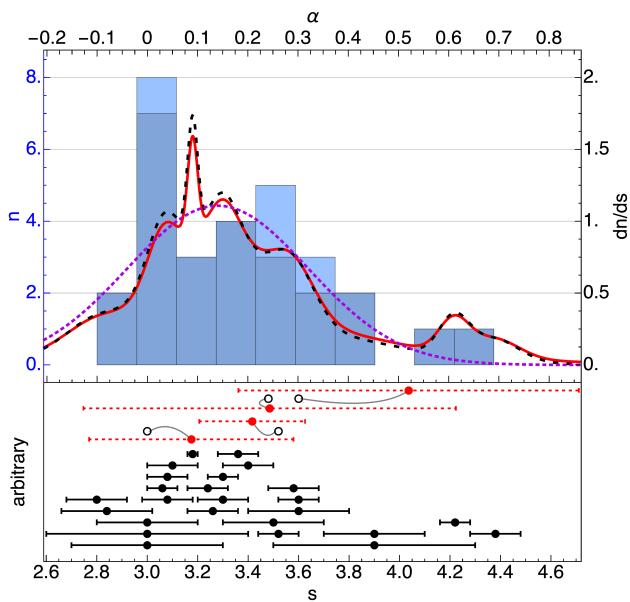}}
    \caption{
    Distribution of radio-based spectra among the 29 PWNe in our sample (see Table \ref{pwn_table}).
    The photon ($\alpha$, top axis) and inferred electron momentum ($s$, bottom axis) spectral indices are shown as raw data (bottom panel) and as reconstructed distributions (top panel: histograms with left axis and spectral density functions with right axis). Data uncertainties are available for most (filled black discs with solid error-bars; dark histogram bars; dashed black curve) but not all (empty black discs; light-blue bars) PWNe.  The latter are supplemented by error propagation-based estimates (filled red discs with dashed error-bars), and are included in the modified spectral density function (solid red curve) and Gaussian fit (dotted purple).}
    \label{fig:histogram}
\end{figure}

Some PWNe show spatial variations in the inferred spectral index.
In such cases, we focus on the spectrum measured locally around the NS, and classify the PWN as core-type, as discussed in \S\ref{sec:classifications}.
Some PWNe show a spectral break, in which case we focus on the spectral index at frequencies lower than the evident break.
The spectral indices of individual PWNe are shown in Fig.~\ref{fig:histogram} (bottom panel), along with the reconstructed  and modelled distributions (top panel).

We exclude from the analysis two PWNe --- N157B and MSH 15-52 --- due to systematic uncertainties, leaving us with the $N_s=29$ PWNe shown in the figure.
In N157B, the spectral index $\alpha=0.1\pm0.2$ is not well differentiated from the spectrum of the surrounding SNR \citep{dickel2004non}.
In MSH 15-52, no distinct radio PWN was detected around the NS \citep{gaensler1999snr}, and the justification for the spectral index $\alpha=0.4$ quoted later \citep[][]{fleishman2007diffusive, greencatalogue} is unclear.

\begin{landscape}
    \begin{table}
        \caption{Radio spectral indices for 31 known PWNe from the literature.
        }
        \centering
        \begin{tabular}{llccclll}
            \hline
            \vspace{0.1cm}
            PWN name & Associated pulsar & $\alpha$ & $s$ & Age [yr] & Alt. Name & Classification & Reference \\
            (1) & (2) & (3) & (4) & (5)& (6)& (7)&(8)\\
            \hline
            B0540-69.3&PSR B0540-69&$0.15\pm0.03$&$3.3\pm0.06$&760&&Integrated&\citet{brantseg2013multi}\\
            J0453-6829&NA&$0.04\pm0.04$&$3.08\pm0.08$&13000&&Integrated&\citet{haberl2012multi}\\
            G000.9+00.1&PSR J1747-2809&$0.18\pm0.04$&$3.36\pm0.08$&3000&&Crab-like&\citet{dubner2008high}\\
            G005.27-00.90&PSR B1757-24&$0.09\pm0.20;(\approx0)^A$&$3.18\pm0.40;(\approx3)^A$&NA&Duck&\textbf{Core-type}; Bow-shock&\citet{frail1991unusual,caswell1987galactic}\\
            G011.2-0.3&PSR J1811-1925&$0.25\pm0.10$&$3.5\pm0.2$&1634&&Crab-like&\citet{tam2002spectral,roberts2003pulsar}\\
            G016.73+0.08&NA&$0.15\pm0.05$&$3.3\pm0.1$&1500&&Integrated&\citet{helfand1989prevalence}\\
            G021.5-0.9&PSR J1833-1034&$-0.08\pm0.09^{\ddagger}$&$2.84\pm0.18$&1000&&Crab-like&\citet{Bietenholz2008,camilo2006psr}\\
            G029.7-0.3&PSR J1846-0258&$0.20\pm0.05$&$3.4\pm0.1$&480&Kes 75&Integrated&\citet{salter198984}\\
            G034.7-00.4&PSR B1853+01&$0.12\pm0.04$&$3.24\pm0.08$&NA&W44&Integrated&\citet{frail1996pulsar}\\
            G054.1+0.3&PSR J1930+1852&$0.13\pm0.05$&$3.26\pm0.10$&NA&&Crab-like&\citet{velusamy1988g54}\\
            G065.7+1.2&NA&$0.45\pm0.20$&$3.9\pm0.4$&20000&DA 495&Integrated&\citet{kothes2008495}\\
            G069.0+02.7&PSR B1951+32&$0.0\pm0.1$&$3.0\pm0.2$&NA&CTB80&\textbf{Core-type}&\citet{castelletti2005multi}\\
            G074.9+1.2&NA&$0.21\pm0.10;(0.26)^A$&$3.42\pm0.21;(3.52)^A$&10000&CTB 87&Integrated&\citet{morsi198732,matheson2013x}\\
            G076.9+1.0&PSR J2022+3842&$0.61\pm0.03$&$4.22\pm0.06$&5000&&Integrated&\citet{marthi2011central,landecker1993g76}\\
            G106.65+2.96&PSR J2229+6114&$0.00\pm0.15$&$3.0\pm0.3$&3900&&\textbf{Core-type}&\citet{kothes2006boomerang}\\
            G130.7+3.1&PSR J0205+6449&$0.09\pm0.01$&$3.18\pm0.02$&839&3C58&Crab-like&\citet{Kothes2017}\\
            G141.2+5.0&NA &$0.69\pm0.05$&$4.38\pm0.10$&NA&&Integrated&\citet{kothes2014g141}\\
            G184.6-5.8&PSR B0531+21&$0.30\pm0.04$&$3.60\pm0.08$&966&Crab&Crab-like&\citet{bietenholz1997radio}\\
            G189.1+3.0&NA&$0.04\pm0.05$&$3.08\pm0.10$&30000&IC443&\textbf{Core-type}; Bow-shock&\citet{castelletti2011high,olbert2001bow}\\
            G263.9-3.3&PSR B0833-45&$-0.10\pm0.06$&$2.80\pm0.12$&NA&Vela X&\textbf{Core-type}&\citet{hales2004vela}\\
            G283.1-0.59&PSR J1015-5719&$0.0\pm0.2$&$3.0\pm0.4$&NA&&\textbf{Core-type}; Bow-shock&\citet{ng2017discovery}\\
            G291.0-0.1&NA&$0.29\pm0.05$&$3.58\pm0.10$&1300&MSH 11-62&Crab-like&\citet{slane2012broadband}\\
            G292.0+1.8&PSR J1124-5916&$0.05\pm0.05$&$3.1\pm0.1$&1600&MSH 11-54&Integrated&\citet{gaensler2003multifrequency}\\
            G315.78-0.23&PSR J1437-5959&$0.45\pm0.10$&$3.9\pm0.2$&19000&Frying Pan&Integrated; Bow-shock&\citet{ng2012extreme}\\
            G319.9-0.7&PSR J1509-5850&$0.26\pm0.04$&$3.52\pm0.08$&NA&&Crab-like; Bow-shock&\citet{ng2010radio}\\
            G327.1-1.1&NA&$0.3\pm0.1$&$3.6\pm0.2$&17000&Snail&Integrated; Bow-shock&\citet{ma2016radio}\\
            G328.4+0.2&NA&$0.03\pm0.03$&$3.06\pm0.06$&10000&&\textbf{Core-type}&\citet{gelfand2007radio}\\
            G341.2+0.9&PSR B1643-43&$0.24\pm0.37;(0.24)^A$&$3.49\pm0.74;(3.48)^A$&NA&&Integrated&\citet{giacani2001pulsar}\\
            G343.1-2.3&PSR B1706-44&$0.52\pm0.34;(0.3)^A$&$4.04\pm0.67;(3.6)^A$&5000&&Integrated&\citet{giacani2001pulsar}\\
            B0538-691&PSR J0537-6910&$0.1\pm0.2^B$&$3.2\pm0.4^B$&NA&N157B&NA&\citet{dickel2004non}\\
            G320.4-1.2&PSR B1509-58&$0.4^B$&$3.8^B$&1835&MSH 15-52&NA&\citet{fleishman2007diffusive}\\
            \hline
        \end{tabular}
        \label{pwn_table}
        \vspace{0.2cm}
        \begin{tablenotes}
            \item Columns: (1) The PWN name; (2) The associated pulsar if detected; (3) The radio spectral index; (4) The inferred particle spectral index; (5) The age of the PWN; (6) Alternative name if relevant; (7) The PWN classification; (8) References.
            Each PWN is classified as either (i) Core-type (in bold), where the spectral index varies across the PWN;
            (ii) Crab-like, where the spectral index is uniform across the entire PWN;
            or
            (iii) Integrated, where it is unknown if the spectrum varies spatially.
            The spectra pertain to a compact region around the \termshock{} for core-type PWNe (bold), and are integrated over the entire PWN for other classes.
            PWN showing a bow shock as they move through the ejecta are labeled as such.
            Spectral indices marked by ($^A$) in columns in columns (3) and (4) lack reported uncertainties. For these, we show the our estimations of these values, and the reported spectrum in brackets. Spectral indices marked with ($^B$) are excluded from all analyses due to systematic uncertainties.~$^{\ddagger}$This uncertainty represents a combination of $\alpha=-0.08_{-0.06}^{+0.09}$ systematic and $\Delta \alpha=\pm0.1$ formal uncertainty.
        \end{tablenotes}
    \end{table}
\end{landscape}

Uncertainty estimates, available for most spectral indices in our sample (filled black discs with solid error-bars, dark-blue bars, and dashed curve), are interpreted as the $1\sigma$ confidence interval of a normal distribution.
For four of the 29 PWNe (empty discs, light-blue bars), no spectral index uncertainty was reported.
We estimate the spectral indices of these PWNe (filled red discs with dashed error-bars) based on the reported flux values, evaluating the uncertainty using error propagation, or through least-squares minimization when possible; the reported values all lie within the $1\sigma$ confidence interval of our estimates.

We reconstruct the spectral density function (SDF; solid curve) by summing the normal distributions associated with each of the spectral index estimates in the sample.
The SDF is broad, with long asymmetric tails.
The spectral index typically lies in the range $-0.30\lesssim\alpha\lesssim 0.26$ ($68\%$ containment), and peaks around $\alpha\simeq0.1$.
A small, sharp peak at $\alpha\simeq0.1$ is associated with one PWN \citep[3C58;][]{Kothes2017} of a particularly small reported uncertainty.
The SDFs including (solid curve) and excluding (dashed) the aforementioned four supplemented indices are consistent with each other.

The distribution of spectral indices is also presented as a histogram, which does not incorporate measurement uncertainties.
The histogram is broadly consistent with the SDF, but highlights two features:
(i) an abundance of spectral indices with a central estimate falling in the $\alpha\simeq 0$ bin;
and
(ii) a possible bimodality, with the second mode naively located around $\alpha\simeq 0.25$.
These features are not sensitive to the bin definitions, and in particular persists as the number of bins is varied in the range $6\simeq N_s/5\leq N_b\leq N_s/2\simeq 14$.
Both features are largely washed out in the SDF, after taking into account measurement errors.
Yet, it is interesting to examine if these or other structures may be hidden in the distribution.

\subsection{Approximately normal spectral distribution}
\label{subsec:Model}

Our sample of spectral indices has a sample mean $\mu_s=0.20\pm0.02$, standard deviation $\sigma_s=0.20\pm0.05$, Pearson moment coefficient of skewness $S_s=0.78\pm0.75$, and Pearson moment coefficient of kurtosis $K_s=2.89\pm1.96$.
These moments are consistent with a normal, and possibly somewhat right-skewed, distribution.

We therefore examine a normal distribution as the null hypothesis model for the spectra in our sample.
As the data set is discrete, non-binned, and characterised by widely varying uncertainty estimates, we maximise the likelihood by treating each measurement as a normal distribution, or by combining the Lilliefors test \citep{lilliefors1967kolmogorov} with Monte-Carlo realizations of the uncertainties.
We also consider maximizing the likelihood of the discretised SDF or the likelihood of an uncertainty-weighted bins, resulting in a comparable but slightly harder distribution.

Consider an unbinned, maximal likelihood analysis that incorporates the measurement errors.
The likelihood $\mathcal{L}_j$ of a given measurement $j$ can be estimated by weighing the model by a normalised Gaussian that represents the distribution of possible measurement values given the reported value and uncertainty. The overall likelihood of the model can then be estimated as $\mathcal{L}=\Pi_j \mathcal{L}_j$, which can be maximised to find its best-fitting parameters. For a normal distribution of spectral indices, the maximal likelihood is obtained for $\alpha\simeq0.19\pm0.18$, consistent with the sample mean and standard deviation.

Similar results, including also a goodness-of-fit estimate, are derived by combining the Lilliefors test with Monte-Carlo simulations of the statistical uncertainties.
The Lilliefors test is a variant of the Kolmogorov-Smirnov (KS) test, useful when the null hypothesis is a normal distribution with unknown parameters, determined from the population mean and variance. As in the KS test, the test-statistic $D_L$ is defined as the maximal offset between the empirical distribution function of the data, and the cumulative distribution function of the model.

When naively ignoring measurement errors and using (henceforth) all originally reported 29 spectral indices in our sample, this test indicates a p-value of $0.15$, so a normal distribution with $\alpha=0.20\pm0.20$ cannot be rejected at any plausible confidence level.
To account for the errors, we use Monte-Carlo simulations, applying the Lilliefors test separately to each realization. Adopting the median value of either $D_L$ or the p-value yields $0.14$ for the latter, similar to the naive estimate.
We conclude that the aforementioned normal distribution cannot be rejected based on these tests.

Another option is to fit a Gaussian model to the SDF, which already accounts for the uncertainty errors.
This procedure leads to a similar but slightly harder spectral distribution, with mean $\bar{\alpha}\simeq0.14\pm0.01$ and standard deviation $\sigma(\alpha)\simeq0.18\pm0.01$, converged as the number of discretised points exceeds $\sim60$ and tends to infinity.
This fit is shown in the top panel of Fig.~\ref{fig:histogram} (dotted purple curve).
As we argue below, the spectrum is slightly harder using this procedure because it is less sensitive to the two $\alpha\simeq 0.6$ outliers.

Similar conclusions are found from a binned analysis, but its results are somewhat sensitive to the details of the binning procedure.
The probability $p_{jk}$ that PWN $j$ lies in spectral bin $k$ is given by the integral within the bin of the normal distribution representing the reported spectral index $\alpha_j$ and its uncertainty.
This assigns bin $k$ with value $\sum_jp_{jk}$ and variance $\sum_jp_{jk}(1-p_{jk})$.
The best-fit normal distribution typically shows $\bar{\alpha}\simeq 0.15$ and $\sigma(\alpha)\simeq0.18$, similar to the SDF analysis, with $\chi_\nu^2\simeq0.6$ chi-squared per degree of freedom corresponding to a p-value of $\leq0.8$.
Here we restrict the analysis to $7\leq N_b\leq13$ bins in the range $-0.28<\alpha<0.84$, which contains the $3\sigma$ confidence intervals of all data points; the results show little dependence upon $N_b$ in this range.

Next, with the null hypothesis of a normal distribution, consider the possibility of an underlying bimodal distribution.
We use the likelihood ratio test, and the resulting TS-test statistic, to examine at what significance level can one rule out the null hypothesis (subscript $0$) in favour of a bimodal distribution (subscript 'bi').
Specifically, we choose a bimodal distribution consisting of the superposition of two Gaussians, and compute $\TS\equiv-2\log\mathcal{L}_{0}/\mathcal{L}_{\mbox{\tiny bi}}$.
This statistics follows a $\chi^2$ distribution with $\nu=\nu_{\mbox{\tiny bi}}-\nu_0$ degrees of freedom, up to order $N_s^{-1/2}$ corrections \citep{wilks1938large}.

The results marginally favor a bimodal distribution, with an $\alpha_2\simeq 0.6$ peak accounting for the two aforementioned soft outliers.
The unbinned, maximal likelihood analysis favors comparable contributions from an $\alpha_1=0.14\pm0.10$ Gaussian and from a sharp (narrower than the measurement uncertainties), $\alpha_2\simeq 0.63$ peak, with $\TS\simeq 11.4$ corresponding to the $2.6\sigma$ confidence level.
The uncertainty-weighted binned analysis weakly favours a bimodal distribution with a main $\alpha\simeq0.14\pm0.16$ Gaussian and a low amplitude, $\alpha\simeq0.64\pm0.09$ peak,
but only at low, $1.3\sigma$--$1.6\sigma$ confidence levels for $7\leq N_b\leq13$.

The above results suggest that the two soft, $\alpha\sim0.6$ outlier PWNe, G076.9+1.0 and G141.2+5.0, are distinguished from the main distribution.
Excluding them leaves a sample mean $0.16\pm0.03$ and standard deviation $0.16\pm0.06$, with $S_s=0.47\pm1.50$ and $K_s=2.48\pm4.24$.
The corresponding Lilliefors test yields a high, $\simeq0.8$ p-value, and the different analysis variants now roughly agree on a single Gaussian model with $\bar{\alpha}\simeq0.14$ and $\sigma(\alpha)\simeq0.13$, similar to the broader peak from the bimodal models.
We conclude that the remaining sample is highly consistent with a normal distribution.
It is interesting to examine any peculiarities in the two outlier PWNe.

The PWN G076.9+1.0 shows central X-ray diffuse emission and two distinct radio lobes, similar to the morphology of DA495 \citep{marthi2011central}.
Its spectrum, $\alpha=0.61\pm0.03$, when integrated over the entire nebula, spreading over a relatively large physical size of $\sim20\pc$ assuming a distance of $\sim7\mbox{ kpc}$.
PWN G141.2+5.0 shows a similarly soft spectrum, $\alpha=0.69\pm0.05$, integrated over the entire nebula of a physical size of $\sim4\pc$ assuming a distance of $\sim4\mbox{ kpc}$ \citep{kothes2014g141}.
In both cases, spatial variations in the spectral index across the PWN were not reported.
We do not identify special properties of these two outliers that might explain their softer spectra.

\subsection{Spectrum--morphology association}
\label{sec:classifications}

\begin{figure}
\centerline{\epsfxsize=8.4cm \epsfbox{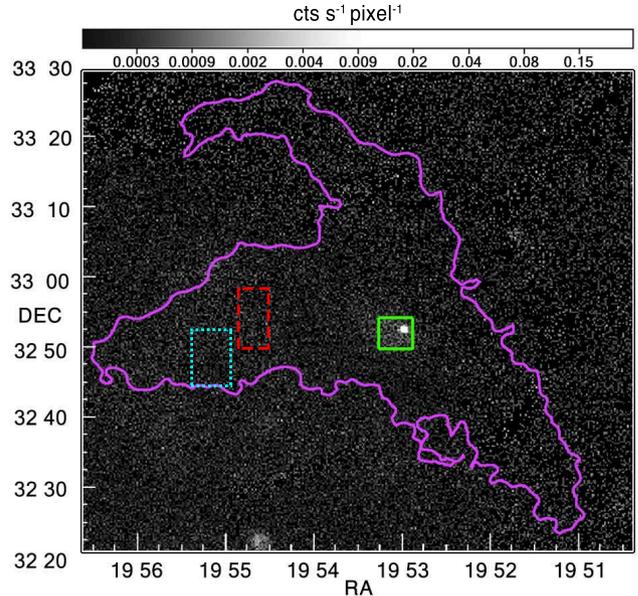}}
    \caption{
    CTB 80 (G069.0+02.7) is an example of a PWN with a spectrally-distinguishable core. The irregular PWN is roughly traced by the radio contour \citep[$30$\text{ mJy/beam} contour of the $1380\MHz$ VLA map;][outer pink contour]{castelletti2005multi}, shown superimposed on the \textit{ROSAT} PSPC image (https:\textbackslash\textbackslash skyview.gsfc.nasa.gov).
    The radio spectrum shows considerable spatial variability; the region (green-solid box) containing PSR B1951+32 defines the core spectrum ($\alpha=0\pm0.1$) \citep{castelletti2005multi} used in the analysis, while the spectrum further away from the core (\eg along the eastern side of the nebula) is as soft as $\alpha\simeq0.6$ (red-dashed box) and $\alpha\simeq0.7$ (cyan-dotted box).
    \label{fig:coreexample}}
\end{figure}

Next, consider the different morphologies of spectral variations across the PWN.
Out of the 29 nebulae in our sample, eight systems show a uniform spectrum across the nebula (\ie are reported to show no variations of the spectrum across the nebula), and are classified as Crab-like in Table \ref{pwn_table}; a prototypical example is the Crab nebula.
Seven systems show variations in the spectrum across the nebula, with a hard core around the NS, and are classified as core-type; a prototypical example is CTB 80 (G069.0+02.7), shown in Fig.~\ref{fig:coreexample}.
In the remaining 14 PWNe, the distribution of the spectrum across the nebula is either unknown or unreported, and only an integrated spectral index is available; these systems are classified as 'integrated'. With future analyses any of the integrated objects could be re-classified as core-type or Crab-like.

\begin{figure}
\centerline{{\epsfxsize=8.4cm \epsfbox{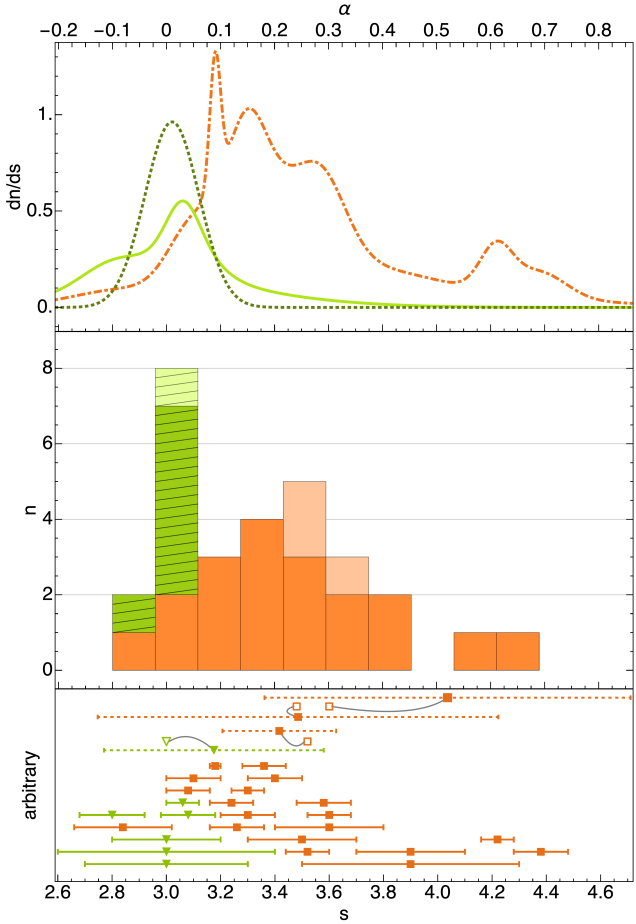}}}
    \caption{
    Same as Fig.~\ref{fig:histogram}, but distinguishing between the seven core-type group PWNe (green triangles; hatched bars) and the rest of the PWNe (non-core-type group; orange squares; solid bars). \textbf{Top panel:} Probability density functions of core-type (solid curve) and non-core-type (dot-dashed curve) spectral indices, normalised with respect to the total number of PWNe, alongside the best-fitted Gaussian model (dotted curve) for core-type PWNe.
    }
    \label{fig:corehistogram}
\end{figure}

Noting a correlation between core-type systems and very hard central spectra around $\alpha\simeq 0$, we examine the possibility that these systems may dominate the low-$\alpha$ part of the distribution.
For this purpose, in Fig.~\ref{fig:corehistogram} we show the spectral distribution among PWNe subdivided between core-type (green triangles, solid curve, and hatched bars) and other (orange squares, dot-dashed curve, and solid bars) classifications.
As the figure shows, the distribution of core-type spectra indeed appears to be distinct from the distribution of other morphological types, and is visibly harder.

The core-type distribution shows a sample mean $0.01\pm0.05$ and standard deviation $0.06\pm0.06$, with higher moments $S_s=-0.55\pm4.72$ and $K_s=2.36\pm10.56$ of substantial uncertainty due to the small sample. In comparison, the combined distribution of the rest of the PWNe shows a sample mean $0.25\pm0.03$ and standard deviation $0.19\pm0.07$, whereas the Crab-like PWNe have a sample mean $0.18\pm0.02$ and standard deviation $0.13\pm0.05$.
Both of these sub-samples appear to have a significantly softer distributions than that of core-type PWNe.

In order to quantify the significance of the distinction between core-type and other PWNe systems, we perform several statistical tests.
These tests are used to compare the core-type PWNe both with all other (Crab and integrated) systems, and specifically with the Crab-like PWNe; note that the integrated systems may include some yet unidentified core-type PWNe. We test the sub-samples both with and without measurement uncertainties; the latter may be relevant if the quoted errors are overestimated, or carry systematic errors common to the sample.

First consider applying the student t-test to examine if core-type and other PWNe have the same mean spectral index. This null hypothesis is rejected at the $\sim3.0\sigma$ level, statistically indicating a harder mean spectrum of the core-type
sub-sample.
Here, we again incorporate uncertainties using the median test-statistics of a large number of Monte-Carlo realizations.
If, instead, measurement uncertainties are ignored, the rejection significance level increases to $\sim4.4\sigma$.

Applying the same test to compare core-type PWNe with only Crab-like systems, while accounting (ignoring) measurement uncertainties, rejects the null hypothesis of equal means at the $\sim2.1\sigma$ ($\sim2.7\sigma$) significance level, lower than above at least in part due to the smaller sample.

Next, we apply the KS test to examine if the core-type spectra may be drawn from the same distribution as other PWNe.
Comparing core-type PWNe to other systems, the null hypothesis of identical underlying distributions is rejected at the $\sim2.7\sigma$
($3.7\sigma$) level when including (neglecting) measurement uncertainties.
Similarly comparing core-type PWNe only with Crab-like systems, rejects the null hypothesis at the $\sim2.4\sigma$ ($3.1\sigma$) level.

These tests support the apparent separation seen in Fig.~\ref{fig:corehistogram}, between core-type PWNe with a hard, $\alpha\simeq 0$ spectrum, and the other, somewhat softer, $\alpha\sim 0.2$ PWN systems.
The distributions differ at the $\sim 3\sigma$ confidence level, and perhaps even at the $\sim4\sigma$ level if the statistical uncertainties are inflated or in part compensated by systematics.
As core-type (seven PWNe) and other (8 and 14 PWNe) sub-samples are small, more data are needed to securely establish or rule out this observation.

Focusing on the core-type PWNe sub-sample, we find it consistent with a normal spectral distribution.
This is seen from the Lilliefors test, giving a p-value $\sim0.40$ (0.06) when including (neglecting) measurement uncertainties.
The Monte-Carlo realizations indicate a mean $\alpha\sim0.01$, with a standard deviation $\sigma(\alpha)\sim0.05$, shown in Fig.~\ref{fig:corehistogram} as a dotted dark-green curve.

\subsection{Implications of $\alpha\simeq 0$}
\label{subsec:Implications}

Let us consider the possibility that core-type PWNe indeed constitute a distinct population, with an $\alpha=0.01\pm0.06$ spectrum significantly harder than in other PWNe.
It is interesting to ask what properties of these PWNe may be unique and associated with their harder spectra, what may be special about the exceptional PWN that shows a similarly hard spectrum in spite of a Crab-like classification, and what are the possible physical implications of such a hard spectrum.

As Table \ref{pwn_table} shows, bow-shock morphologies are more prevalent among core-type PWNe, found in three out of the seven core-type PWNe, in comparison to three out of the 22 PWNe of other classifications.
Core-type PWNe are also on average somewhat older than PWNe in other classes, with a median age $\sim7\kyr$ vs. $\sim1.6\kyr$ excluding bow shocks, and $\sim10\kyr$ vs. $\sim1.6\kyr$ when including bow shocks.
These properties suggest that an older, quenched or stripped PWN may be more susceptible for the emergence of a hard spectrum;
for instance, core-type PWNe G328.4+0.2 \citep{gelfand2007radio} and Vela-X \citep{hales2004vela} have probably already interacted with the SNR reverse shock.

It is worth mentioning the exceptional case of G021.5-0.9 \citep{Bietenholz2008}, classified as a Crab-type PWN and yet showing
a very hard, average $\alpha=-0.08^{+0.09}_{-0.06}$ spectrum, where the reported uncertainty is predominantly systematic.
This is a young ($\sim1\kyr$) PWN, with no evidence for an interaction with the reverse shock or for a bow shock.
The spectral map is based on only two frequencies, $1.5\GHz$ and $4.9\GHz$.
RMS fluctuations $\Delta\alpha\simeq 0.14$ were reported across the nebula, with a formal uncertainty of $\Delta\alpha=0.1$ on average.
While the PWN was reported as spectrally uniform, the spectral map appears to show variations in $\alpha$, featuring a $\Delta\alpha\sim0.1$ hardening in the region near the NS.
If this PWN is reclassified as core-type with $\alpha=-0.18$, the distribution of this sub-sample becomes $\alpha=-0.02\pm0.09$, more significantly distinguished from the remaining (or Crab-like) PWNe, for example at a student-t $\sim4.6\sigma$ ($\sim3.8\sigma$) confidence level.

The harder spectra of the core-type group appear to be directly associated with the acceleration of electrons, whereas the highly uniform spectra of Crab-type PWNe could be affected by evolutionary effects as the
cooling time (\ref{eq:sych_cool})
is long.
Interestingly, an $\alpha\simeq 0$ spectral index, as inferred for core-type PWNe, plays an important role in Fermi acceleration and its manifestation in the DSA mechanism.
Here, $\alpha=0$ is the hardest possible spectrum, for any equation of state and in any dimension, corresponding to a vanishing particle escape probability or an infinite energy gain per cycle.
Therefore, unless $\alpha=0$ is shown to be a natural outcome of some competing acceleration mechanism, the results suggest that (i) DSA might be responsible for the acceleration of the radio-emitting electrons; and (ii) DSA nearly saturates the hard spectral limit for the physical conditions around the PWN termination shock.
It is unclear why DSA should approach the $\alpha=0$ limit in PWNe, unlike in other relativistic shock systems, and if such a hard spectrum is even viable in the small-angle scattering limit; we consider these issues in the following sections.

\section{DSA analysis}
\label{sec:method}

We model the termination shock as a planar, $\gamma_u=10^4$ shock with the J\"uttner--Synge equation of state \citep{juttner1911maxwellsche, synge1957relativistic}, such that $\beta_u-1\simeq -5.000\times 10^{-9}$ and $\beta_d-1/3\simeq -6.667\times10^{-9}$.
While it is unclear if linear DSA is relevant to particle acceleration in the termination shock, it is interesting to ask if it could lead to the hard spectrum observed.
The PDF and spectrum of the accelerated particles are derived semi-analytically using an expansion in moments of the angular distribution \citep[following][]{Keshet06}.

In the following, \S\ref{sec:dsa_setup} presents the setup and formalism of the relativistic, small-angle scattering, DSA model.
The moment solution is reviewed and implemented in \S\ref{sec:moments_method}.
In \S\ref{sec:energain}, we quantify the acceleration process in terms of the return probability and the energy gain in a Fermi cycle, and show how both can be accurately extracted as useful diagnostics directly from the PDF.

\subsection{Setup}
\label{sec:dsa_setup}

Consider an infinite planar shock front located at shock-frame coordinate $z=0$, with flow in the positive $z$ direction.
We study the acceleration of relativistic particles of momenta $p$ much larger than any characteristic scale in the system, due to repeated small-angle scatterings assumed to be elastic in the fluid frame (denoted by a tilde).
Here, the evolution of the particle direction $\tilde{\mu}=\cos(\mathbf{\tilde{p}}\cdot\hat{z}/\tilde{p})$ is approximately diffusive.

Adopting some prescribed angular diffusion-function (DF) $\tilde{D}_{\mu \mu}$ on each side of the shock, the steady-state PDF $f$ satisfies the transport equation \citep[\eg][]{kirk1987acceleration},
\begin{equation}
\label{transport_a}
    \gamma_i(\tilde{\mu}_i+\beta_i)\frac{\partial{f}(\tilde{\mu}_i,\tilde{p}_i,z)}{\partial{z}} = \frac{\partial}{\partial{\tilde{\mu}_i}}\left[\tilde{D}^{(i)}_{\mu\mu}(\tilde{\mu}_i,\tilde{p}_i,z)\frac{\partial{f}}{\partial{\tilde{\mu}_i}}\right]\fin
\end{equation}
Here $\gamma_i\equiv(1-\beta_i^2)^{-1/2}$ is the shock-frame bulk Lorentz factor of the fluid, and the upstream/downstream index $i\in\{u,d\}$ is used only when necessary.
Variables without a tilde are measured in the shock-frame, so $f(\tilde{\mu}_i,\tilde{p}_i,z)$ is the Lorentz-invariant density in a mixed-frame phase space.

The PDF is continuous across the shock front,
$f_u(\tilde{\mu}_u,\tilde{p}_u,z=0)=f_d(\tilde{\mu}_d,\tilde{p}_d,z=0)$, where upstream and downstream parameters are related by a Lorentz boost of normalised velocity $\beta_r=(\beta_u-\beta_d)/(1-\beta_u\beta_d)$, namely $\tilde{p}_d=\gamma_r\tilde{p}_u(1+\beta_r\tilde{\mu}_u)$ and $\tilde{\mu}_d=(\tilde{\mu}_u+\beta_r)/(1+\beta_r\tilde{\mu}_u)$.
The PDF is fixed by the upstream boundary condition, requiring the absence of accelerated particles far upstream, $f_u(z\rightarrow-\infty)=0$.
In the absence of a relevant momentum scale, a power-law spectrum develops \citep[\eg][]{Bednarz_ostrowski1996,achterberg2001particle}, so we may separate the PDF in the form $f_i(\tilde{\mu}_i,\tilde{p}_i,z)\equiv\tilde{q}_i(\tilde{\mu},\tilde{\tau})\tilde{p}_i^{-s}$.
The PDF naturally becomes isotropic in the fluid frame far downstream, $f_d(z\rightarrow\infty)=\qdinf\tilde{p}_u^{-s}$, where $\qdinf>0$ is a constant.

Under certain assumptions \citep[\eg][]{Kirk_2000}, the angular dependence of the DF can be separated out, $\tilde{D}_{\mu\mu}=(1-\tilde{\mu}^2)\tilde{D}(\tilde{\mu})\tilde{D}_2(\tilde{p},z)$.
While this separation is not in general justified \citep[\eg][]{Katz_etal_07,nagar2019diffusive}, we adopt it here for simplicity.
The spatial and momentum dependencies can then be absorbed by re-scaling the spatial coordinate as
\begin{equation}
    \label{eq:tau_z}
    \tilde{\tau}\equiv\gamma_i^{-1}\int_{0}^{z}\tilde{D}_2(\tilde{p},z^{\prime})dz^{\prime}\coma
\end{equation}
leading to a dimensionless transport equation for the reduced PDF $q$,
\begin{equation}
    \label{transport_b}
    (\tilde{\mu}+\beta_i)\partial_{\tilde{\tau}}{\tilde{q}_i}(\tilde{\mu}_i,{\tilde{\tau}})=\partial
    _{\tilde{\mu}}\left[(1-\tilde{\mu}^2)\tilde{D}(\tilde{\mu})\partial_{\tilde{\mu}_i}\tilde{q}_i\right] \fin
\end{equation}

\begin{figure}
    \centering
    \begin{tikzpicture}
        \raggedleft
        \node[anchor=south west,inner sep=0] at (-0.3,0-1.4){
        \centerline{\epsfxsize=8.4cm  \epsfbox{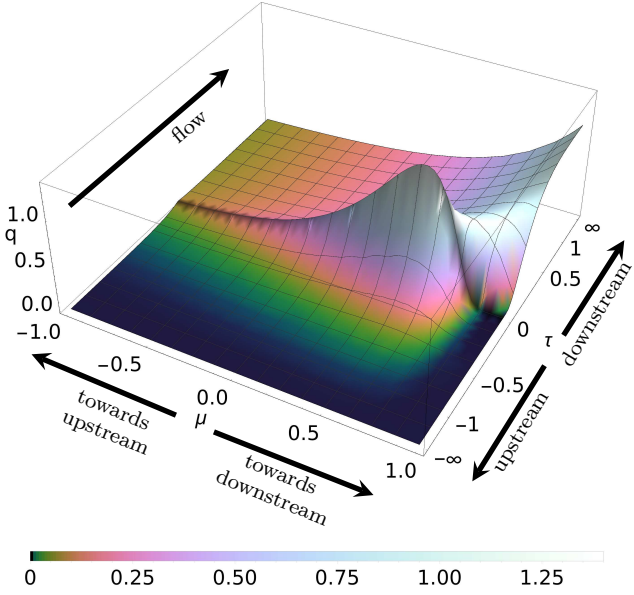}}};
        \draw[line width=2pt,black,-stealth](7.15,1.7)--(6.2,0.18);
        \draw[line width=2pt,black,-stealth](7.4,2.1)--(8.2,3.35);
        \node[label={[label distance=0.5cm,text depth=-1ex,rotate=59]right:upstream}] at (6.1,-0.2) {};
        \node[label={[label distance=0.5cm,text depth=-1ex,rotate=59]right:downstream}] at (7.0,1.2) {};
        \draw[line width=2pt,black,-stealth](0.9,3.8)--(3,5.6);
        \node[label={[label distance=0.5cm,text depth=-1ex,rotate=40]right:flow}] at (1.7,4.3) {};
        \draw[line width=2pt,black,-stealth](2.8,0.83)--(4.9,0);
        \node[label={[label distance=0.5cm,text depth=-1ex,rotate=-22]right:towards}] at (2.55,0.7) {};
         \node[label={[label distance=0.5cm,text depth=-1ex,rotate=-22]right:downstream}] at (2.18,0.5) {};
        \draw[line width=2pt,black,-stealth](2.3,1.05)--(0.4,1.85);
        \node[label={[label distance=0.5cm,text depth=-2ex,rotate=-22]right:towards}] at (0.35,1.55) {};
         \node[label={[label distance=0.5cm,text depth=-2ex,rotate=-22]right:upstream}] at (0.15,1.28) {};
    \end{tikzpicture}
    \caption{
    Reduced shock-frame PDF $q$, shown \citep[height and cubelix,][coluorbar]{cubehelix}
    in the full phase space of particle direction $\mu$ vs. optical depth $\tau$ from the shock,
    for isotropic diffusion around (henceforth) our nominal, ultra-relativistic shock of $\gamma_u=10^4$ with a J\"uttner--Synge equation of state.
    The PDF is calculated
    using $N=35$ Lagrange moments  (see \S\ref{sec:moments_method}), and normalized such that $\int_{-1}^{+1}q(\mu,\tau=0)d\mu=1$.
    The displacement $\tau$ from the shock front is compactified as $\tanh\tau$.
    }
    \label{fig:angualr_spatial_iso}
\end{figure}

It is advantageous for our purposes to write all variables in the shock frame, leading to the reduced transport equation \citep{Keshet06}
\begin{equation}
    \label{transport_c}
    \partial_{\tau}q(\mu,\tau)=\frac{\partial_{\mu}\left\{(1-\mu^2)D(\mu)\partial_{\mu}\left[(1-\beta\mu)^s q\right]\right\}}{(1-\beta\mu)^{s-3}\mu}\coma
\end{equation}
with $q(\mu,\tau)=\tilde{q}_i(\tilde{\mu}_i,\tilde{\tau})(\tilde{p}_i/p)^{-s}=\tilde{q}_i(\tilde{\mu}_i,\tilde{\tau})\gamma_i^{s}(1+\beta_i\tilde{\mu}_i)^{s}$, $D(\mu)=\tilde{D}(\tilde{\mu})$, and $\tau=\gamma^4\tilde{\tau}$.
The phase space spanned by $\mu$ and $\tau$ is demonstrated (with labeled arrows) in Fig.~\ref{fig:angualr_spatial_iso}, which also shows the reduced-PDF $q(\mu,\tau)$ for the simple case of isotropic diffusion.

\subsection{Solution using moments}
\label{sec:moments_method}

We use the moments method of \citet{Keshet06}. Here, the shock-frame transport equation (\ref{transport_c}) is multiplied by weights $f_n(\mu)$ for $n=0,1,2,\ldots N$, and integrated to yield coupled first-order ordinary differential equations for the spatial evolution of the corresponding scalar moments $F_n\equiv \int f_n q\,d\mu$.
Requiring the continuity of each moment across the shock reduces the problem to a set of $N+1$ algebraic equations for the moment coefficients.
Moments that diverge downstream or do not vanish upstream are discarded.
The remaining equations can be formally solved and restated as a transcendental equation for $s$; the solution converges rapidly with $N$ for an appropriate choice of weights.

We adopt Legendre weights, $f_n(\mu)=(n+1/2)^{1/2}P_n(\mu)$, where $P_n(\mu)$ is the Legendre polynomial of order $n$.
The moment $F_n$ then reduces to the $n$'th-coefficient of the Legendre series
of $q$, and the solution for $s$ converges rapidly \citep{Keshet06}.
For such orthogonal polynomial weights, it is necessary to separate each surviving moment into a uniform component and a component that decays exponentially away from the shock.
For given shock and DF, we compute $s$ and the PDF for increasing choices of $N$, using an increasingly high precision, up to a high order $N_{max}$ ensuring the convergence of $s$ to at least four decimal places.
We estimate $s$ and its uncertainty $\Delta s$ using Richardson extrapolation to $N\to\infty$ and its difference from the $N_{max}$ estimate.
For a more comprehensive discussion of the moments method, see Arad \& Keshet (in preparation).

\begin{figure}
\centerline{\epsfxsize=8.4cm \epsfbox{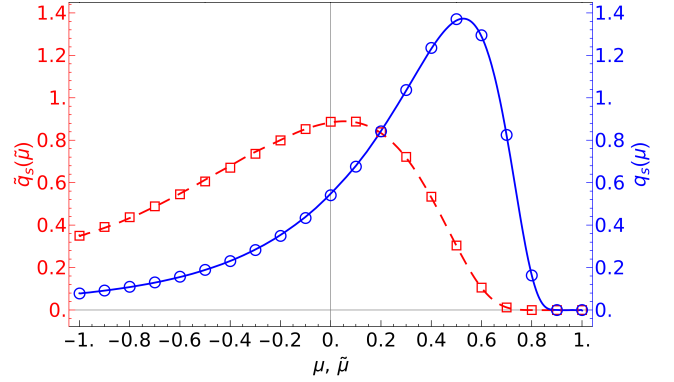}}
    \caption{
    Shock-front angular distributions $q_s(\mu)$ in the shock frame (solid blue curve; right axis) and $\tilde{q}_s(\tilde{\mu}_d)$ in the downstream frame (dashed red; left axis), for the nominal ultra-relativistic shock with isotropic diffusion shown in Fig.~\ref{fig:angualr_spatial_iso},
    and with the same integral normalization in the respective frame.
    The first upstream eigenfunction \citep{Kirk_2000} is shown, for comparison, in both frames (blue circles in shock frame; red squares downstream).
    }
    \label{fig:iso_dis}
\end{figure}

Consider first the simple case of isotropic diffusion both upstream and downstream, $D(\mu)=\const$.
Here, $N=15$ moments are sufficient to reach $\mysiso=4.2270$, converged to four decimal digits, in agreement with previous computations but pushing deeper into the ultra-relativistic shock limit.
This setup is labeled \isoc{}, and is summarised in Tables \ref{diff_table_up} and \ref{diff_table_down}, along with other scenarios discussed below.
For isotropic diffusion we explore higher moments, showing that $N=35$ moments are needed to reach ten-digit precision.
This higher accuracy result is used to produce the reduced PDF shown in Fig.~\ref{fig:angualr_spatial_iso} in the entire compactified phase space, and the shock angular distribution, $q_s(\mu)\equiv q(\mu,\tau=0)$, shown in Fig.~\ref{fig:iso_dis}.
The latter shows the distribution in both shock and downstream frames, along with an approximation for the first upstream eigenfunction \citep{Kirk_2000}, $\tilde{q}_s(\tilde{\mu}_u)\sim\exp[-(1+\tilde{\mu}_u)/(1-\beta_u)]$.

\subsection{Return probability and energy gain}
\label{sec:energain}

By considering Fermi cycles, the spectral index can be expressed in the form \citep{fermi1949origin,Bell_1978}
\begin{equation}
\label{eq:Fermi_relation}
    s=3-\frac{\ln \myPret}{\ln\energain},
\end{equation}
where $P_{\text{ret}}$ is the probability for an accelerated particle crossing downstream to return upstream, $\energain\equiv\llangle E_{j+1}/E_j\rrangle$ is the flux-averaged energy gain per cycle, and $\llangle\ldots\rrangle$ designates flux averaging.
The quantities $\myPret$ and $\energain$ are useful diagnostics of the acceleration process.
They can be evaluated once the PDF has been determined, although $\energain$ depends somewhat on correlations between the directions in which particles cross the shock back and forth.

\subsubsection{Return probability}

By definition, $\myPret$ is the ratio between the downstream-frame magnitudes of the particle fluxes crossing the shock back towards the upstream (subscript $-$), $|j_-|=-j_->0$, and forward towards the downstream (subscript $+$), $j_+>0$,
\begin{equation}
    \label{pret}
    \myPret \MyEqual_d \frac{-j_-}{j_+} =\frac{-j_-^{(d)}}{j_+^{(d)}}\coma
\end{equation}
where the equality subscript $i$ signifies that the subsequent expression is evaluated in the $i$ fluid frame.
Here, we define the shock-front flux element
\begin{equation}
    dj\equiv(\tilde{\mu}+\beta)\tilde{q}(\tilde{\mu},\tau=0)d\tilde{\mu}\coma
    \label{eq:flux_element}
\end{equation}
and its integrals over the relevant angular range, namely
\begin{equation}
j_-\equiv\int dj_-\equiv \int_{-1}^{-\beta}\frac{dj}{d\tilde{\mu}}d\tilde{\mu}\,,
\end{equation}
\begin{equation}
j_+\equiv\int dj_+\equiv\int_{-\beta}^{1}\frac{dj}{d\tilde{\mu}}d\tilde{\mu}\, ,
\end{equation}
and $j\equiv j_-+j_+$.

Equivalently, we may calculate the escape probability $\myPesc=1-\myPret$ as the downstream-frame ratio between the (conserved) total flux
and the forward flux at the shock,
\begin{equation}
    \label{eq:pesc}
    \myPesc
    \MyEqual_d \frac{j}{j_{+}}
    \MyEqual_d \frac{2\beta\qdinf}{\int_{-\beta}^{1}(\tilde{\mu}+\beta)\tilde{q}(\tilde{\mu},0)d\tilde{\mu}}\, .
\end{equation}
For the angular distribution $q(\mu)$ of the isotropic-diffusion case \isoc{}, we find $\myPret\simeq0.379$.

\subsubsection{Energy gain}

In principle, the energy gain can be computed precisely, and in either fluid frame.
The downstream-cycle gain, $\energain_d$, for a particle crossing the shock towards the upstream at some angle $-1<\tilde{\mu}_{-}^{(d)}<-\beta_d$ and returning downstream with probability $P(\mu_+,\mu_-)$ (omitting the superscript $d$ for brevity) at an angle $\beta_d<\tilde{\mu}_{+}<1$ is
$(1-\beta_r\tilde{\mu}_{-})/(1-\beta_r\tilde{\mu}_{+})$.
Hence, averaging over $\tilde{\mu}_-$ yields
\begin{equation}\label{eq:energygain_d}
    \energain_d \MyEqual_d  \left\llangle\frac{1-\beta_r\tilde{\mu}_{-}}{1-\beta_r\tilde{\mu}_{+}}\right\rrangle \MyEquiv_d \frac{\mathlarger{\int}\frac{1-\beta_r\tilde{\mu}_{-}}{1-\beta_r\tilde{\mu}_{+}} P(\mu_+,\mu_-) dj_{-}dj_{+}}
    {\int P(\mu_+,\mu_-)dj_{-}dj_{+}}\fin
\end{equation}
The equivalent expression for the upstream gain, $\energain_u$, is
\begin{equation}
\label{eq:energygain_u}
    \energain_u \MyEqual_u \left\llangle\frac{1+\beta_r\tilde{\mu}_{+}}{1+\beta_r\tilde{\mu}_{-}}\right\rrangle
    \MyEquiv_u \frac{\mathlarger{\int}\frac{1+\beta_r\tilde{\mu}_{+}}{1+\beta_r\tilde{\mu}_{-}}P(\mu_-,\mu_+)dj_{-}dj_{+}}
    {\int P(\mu_-,\mu_+)dj_{-}dj_{+}}\fin
\end{equation}
One can neglect the correlations between $\mu_-$ and $\mu_+$ by approximating $P$ as a constant, but the resulting $\energain_d$ and $\energain_u$ are generally inaccurate and unequal.

In this approximation, we can compute $\energain_d$ and $\energain_u$ directly from the PDF.
Is it convenient to work in the shock frame, by defining the flux element in fluid $i$ as
\begin{equation}
dJ^{(i)}_{\pm}\equiv (1-\beta_i\mu_{\pm})^{s-3}\mu_{\pm} q(\mu_{\pm})d\mu_{\pm} \coma
\end{equation}
where $-1\leq  \mu_-\leq 0\leq \mu_+\leq 1$, such that
\begin{equation}
   \label{eq:eta_shock}
    \energain_i \simeq \frac{\int\frac{1-\beta_u\mu_-}{1-\beta_u\mu_+}\frac{1-\beta_d\mu_+}{1-\beta_d\mu_-} dJ^{(i)}_{-}dJ^{(i)}_{+}}{\int dJ^{(i)}_{-}dJ^{(i)}_{+}}\fin
\end{equation}
Consider the isotropic diffusion case \isoc{}. Here we find that $\energain_u\simeq2.208$ and $\energain_d\simeq2.488$ indeed differ.
Equation (\ref{eq:Fermi_relation}) suggests, based on the aforementioned values of $s$ and $\myPret$, that $\energain\simeq 2.205$, so $\energain_u$ is approximately correct while $\energain_d$ is off.

We find that this is the case quite generally, for a wide range of DFs.
In \S\ref{sec:hard_spectrum} we consider 23 different choices of the DF.
Eight of these choices include anisotropic diffusion upstream, 12 include anisotropic diffusion downstream, and three feature an anisotropic DF both upstream and downstream (for details, see Tables~\ref{diff_table_up} and \ref{diff_table_down}).
For each of these cases, we calculate $\myPret$ and the energy gain approximations $\energain_d$ and $\energain_u$.
Equation (\ref{eq:Fermi_relation}) now assigns each case with an estimate $s_d$ based on $\energain_d$, and an estimate $s_u$ based on $\energain_u$.

\begin{figure}
\centerline{\epsfxsize=8.4cm \epsfbox{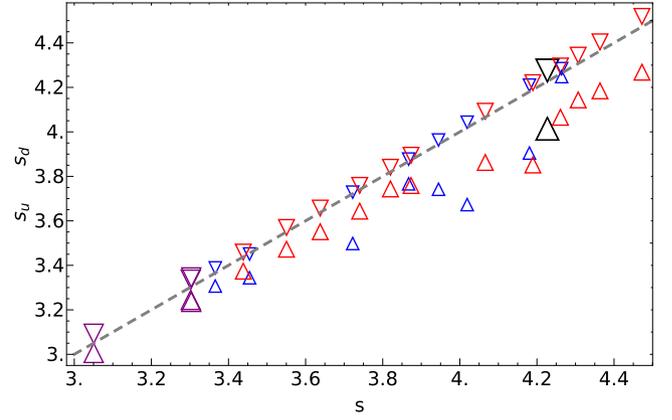}}
    \caption{
    The spectral index $s$ (in the moments method; abscissa) and its approximations (ordinate) $s_u$ (bottom point of down triangles) and $s_d$ (upper point of up triangles) for different choices (see Tables \ref{diff_table_up} and \ref{diff_table_down}) of anisotropic diffusion upstream (small blue markers), downstream (medium red), or both (large purple), as well as for the isotropic diffusion case (largest black).
    The $s_u$ approximation is quite good ($s_u=s$ shown as dashed grey curve), whereas the $s_d$ approximation is not.
    }
    \label{fig:energain_spectrum}
\end{figure}

Figure \ref{fig:energain_spectrum} compares $s_d$ and $s_u$ with the spectral index $s$ inferred from the moments method.
We find that $\energain_u$ provides a good approximation, with an error $|s-s_u|\lesssim0.03$ and typically even an order of magnitude smaller.
In contrast, the $\energain_d$ approximation is rather inaccurate, with $|s-s_d|\lesssim0.3$.
Indeed, after a particle crosses downstream, it has time to diffuse, catch up with the shock, and return upstream in any angle, so the correlations between $\mu_-^{(u)}$ and $\mu_+^{(u)}$ are small and $P_u$ can be approximated as a constant.
In contrast, after a particle crosses upstream, it is swept up by the shock even after a small deflection, generating stronger correlations between $\mu_+^{(d)}$ and $\mu_-^{(d)}$, so neglecting the variations in $P_d$ yields an inaccurate $\energain_d$.
Henceforth, we thus adopt $\energain_u$ as our energy gain estimate.

\section{Hard spectrum conditions}
\label{sec:hard_spectrum}

It is interesting to ask how hard can the DSA spectrum become in the small-angle scattering limit, and what types of DFs can facilitate such a spectrum.
The Fermi cycle Eq.~(\ref{eq:Fermi_relation}) suggests two different routes in which the spectrum may harden with respect to the nominal case of isotropic diffusion: a larger escape probability, or a larger energy gain, both leading at their extreme to the hard limit
\begin{equation} \label{eq:s_limit}
s(\myPret\to 0)=3=s(\energain\to \infty) \,.
\end{equation}

Here, we show how small-angle scattering can approach this limit in both ways, by admitting different changes to the DF.
Thus, anisotropic upstream diffusion can raise $\energain$ substantially, whereas anisotropic downstream diffusion can reach arbitrarily close to $\myPret=1$.
In both cases, the scattering of particles moving away from the shock is suppressed.

We start by examining the effect of anisotropic upstream diffusion in \S\ref{sec:hard_spec_g}, detailing different choices of $D_u$ in Table~\ref{diff_table_up}, with a few special cases labeled, \difcases{U0}, \difcases{U1} \ldots\, highlighted in the text.
In \S\ref{sec:hard_spec_pret}, we study the effect of anisotropic downstream diffusion, with different choices of $D_d$ detailed in Table~\ref{diff_table_down} and special cases labeled \difcases{D0}, \difcases{D1} \ldots\, highlighted in the text.

\subsection{Modified upstream diffusion raising $\energain$}
\label{sec:hard_spec_g}

The $\energain_u$ approximation (\ref{eq:eta_shock}) indicates that for ultra-relativistic shocks with $s<4$, the energy gain can grow very large if the PDF at forward, $\mu\to +1$ directions remains non-negligible, as the numerator integrand is proportional to $(1-\beta_u\mu_+)^{s-4}q_s(\mu_+)$.
Therefore, an angular DF that pushes upstream particles towards $\mu=1$ can substantially raise $\energain$ and thus harden the spectrum.
This conclusion could possibly be inferred from the exact Eq.~(\ref{eq:energygain_d}) for any $s$, but the argument is complicated by the unkown form of $P(\mu_+,\mu_-)$.

Maximizing $\energain_u$ over all possible DFs is computationally challenging, so we demonstrate the behaviour by solving the problem for a selection of representative upstream DFs, keeping the downstream DF isotropic for the moment.
Although the dependence of the spectrum upon the DF is not monotonic \citep{Keshet06}, nor linear \citep{nagar2019diffusive}, taking a linearly increasing (decreasing) upstream DF $D_u(\mu_u)$ does raise (lower) $q_s(\mu\simeq 1)$ and thus $g$, and consequently hardens (softens) the spectrum.
We confirm this behaviour using linear choices of $D_u$, but the effect is quite modest, with the spectral index varying by $|\Delta s|\lesssim 0.05$ \citep{Keshet06}.

\begin{figure}
    \centerline{\epsfxsize=8.4cm \epsfbox{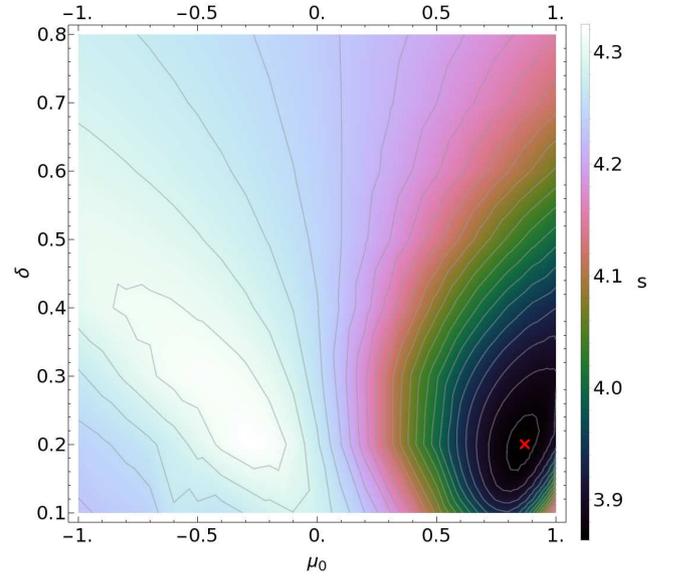}}
    \caption{Spectral index $s$ (colourbar and contours of $\Delta s=0.02$ intervals starting at $s=3.87$) for the nominal shock, as a function of the parameters $\mu_0$ and $\delta$ of the upstream Gaussian DF $D(\mu)=1+100\exp[(\mu-\mu_0)^2/(2\delta^2)]$, with an isotropic DF downstream.
    The hardest case is highlighted (red cross with $s\simeq3.864$ for $\{\mu_0=0.87,\delta=0.20\}$),
    and approximately represented by case \difcases{U2} (see  Table~\ref{diff_table_up}).
    For $\delta>0.1$ ($\delta=0.1$) we use $N=25$ ($N=30$ to $40$) moments.
    }
    \label{fig:upstream_parameters}
\end{figure}

A stronger effect can be obtained by considering changes in the DF that are localised around some $\mu_0$ \citep{Keshet06}.
We thus introduce a DF with a Gaussian enhancement of the form
\begin{equation}
D_G(\mu,h,\mu_0,\delta)\equiv 1+h\exp[-(\mu-\mu_0)^2/2\delta^2] \,,
\end{equation}
and sample $D_u=D_G$ with different parameters.
Figure \ref{fig:upstream_parameters} illustrates the spectrum for the nominal shock with a range of $\mu_0$ and $\delta$, fixing for simplicity $h=100$.
The spectrum is seen to generally harden for large $\mu_0$ and small $\delta$, with the minimum found at $\{\mu_0\simeq0.87, \delta\simeq0.20\}$.
The results depend somewhat on $h$, with the minimal $s$ obtained for different parameters and in general hardening for larger $h$.

Focusing on DFs leading to a hard spectrum, we consider case \difcases{U1} with $\mu_0=0.9$, $\delta=0.2$, and $h=10^6$ upstream, and isotropic diffusion downstream.
This setup yields a very high,
$\energain_u\simeq373$ energy gain, so although the return probability $\myPret\simeq 0.118$ is low, the spectrum $s\simeq3.3657$ is hard.
Such a DF, while radical in its large, factor $\sim10^6$ enhancement of diffusion in the forward direction, demonstrates how small-angle scattering can yield a hard spectrum, probably down to its limit (\ref{eq:s_limit}).
It is numerically difficult to study DFs giving such hard spectra, as the PDF becomes highly concentrated near $\mu=1$ and its resolution requires an increasingly large number of moments.

\begin{figure}
\centerline{\epsfxsize=8.4cm \epsfbox{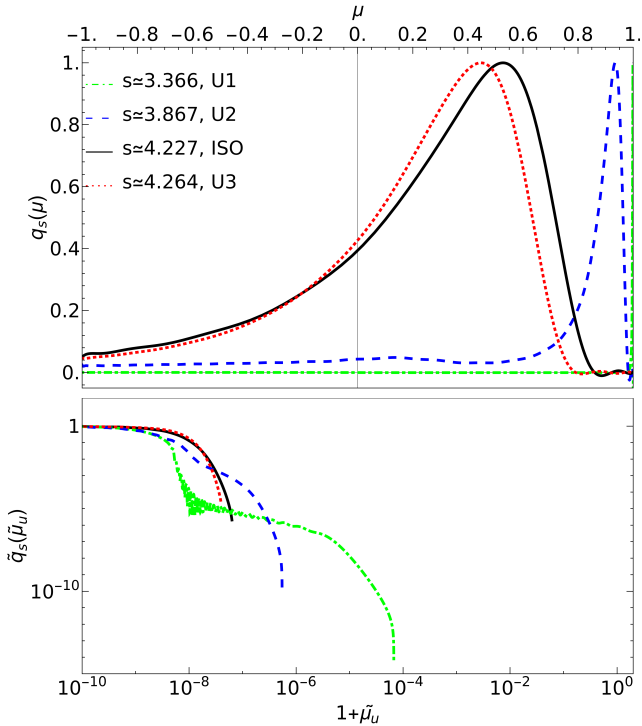}}
   \caption{
    Shock-front angular PDF in the shock (top) and upstream (bottom) frames, for the nominal shock with different spectra (legend) mainly due to different energy gains in upstream DF cases \isoc{} (solid black curve), \difcases{U1} (dot-dashed green), \difcases{U2} (dashed blue) and \difcases{U3} (dotted red).
   The distributions are normalised to unit maximum, $\max[q_s(\mu)]=1$ in the respective reference frame, for ease of comparison. For more details on each case,
   see Table~\ref{diff_table_up}.
   }
    \label{fig:pdf_cases_up}
\end{figure}

The shock-front angular PDF $q_s(\mu)$ is shown in the shock frame for three representative cases (\difcases{U1}, \difcases{U2}, and \difcases{U3}) in the top panel of Fig.~\ref{fig:pdf_cases_up}, alongside case \isoc{}.
As expected, a harder spectrum is associated with an increased fraction of forward-moving particles.
This is most apparent for \difcases{U1} (dot-dashed; blue), in which most particles are highly beamed in the forward, downstream direction.
The figure also illustrates how the mode $\mu_{\mbox{\tiny max}}$ of $q_s(\mu)$ increases monotonically in the forward direction as $\mu_0$ increases and the spectrum becomes harder.

The bottom panel of the figure shows the PDF in the upstream frame.
Although the PDF is beamed in the forward directions in the shock and downstream frames, it retains a large component at fluid-frame angles $\pi-\theta\lesssim \gamma_u^{-1}$.
The fluid-frame DF in the anisotropic cases \difcases{U1}, \difcases{U2},
and \difcases{U3} is approximately constant for $\tilde{\mu}_u>10^{-7}$, vanishing rapidly for smaller $\tilde{\mu}_u$, and so strongly anti-correlated with the PDF.

The spectrum is a non-separable function of upstream and downstream DFs, and scanning the full functional is beyond the scope of this work.
Suffice to point out that the spectra of the above cases can be further hardened by modifying the downstream DF.
For instance, for case \difcases{U0}, combining $D_u$ from \difcases{U1} and $D_d(\mu)=\exp(\mu)$ yields a harder, $s\simeq3.303$ spectrum.
The modified $D_d$ lowers the energy gain (to $g\simeq354$) with respect to \difcases{U1}, but raises the return probability (to $\myPret\sim0.169$), thus lowering $s$.

\begin{figure}
	\centerline{\epsfxsize=8.4cm \epsfbox{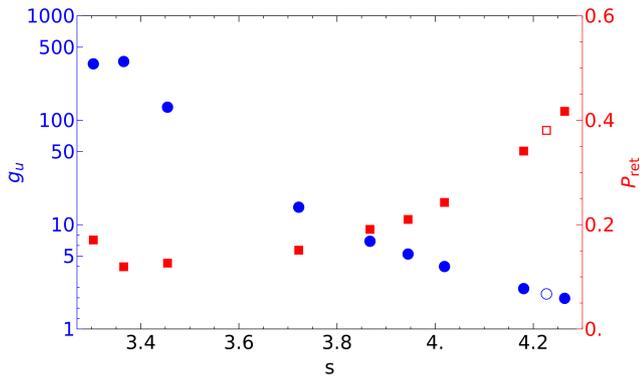}}
	\caption{The energy gain $\energain_u$ (blue circles; left axis) and return probability $\myPret$ (red squares; right axis) as a function of the spectral index for the nine anisotropic upstream diffusion function (see Table~\ref{diff_table_up}) and for the isotropic case (empty markers).
	}
	\label{fig:pret_s_up}
\end{figure}

Figure \ref{fig:pret_s_up} shows the values of $\energain$ (blue disks) and $\myPret$ (red squares) for the different choices of DFs.
The general trend of spectral hardening with an increasing $\energain$ is apparent, with the highest $\energain$ found for \difcases{U1}.
The increasingly forward-peaked upstream DF also leads to a generally decreasing $\myPret$, due to the diminishing fraction of backwards-moving particles at the shock front, but the associated increase in energy gain dominates the behaviour of the spectrum.
The joint effect of modifying both $D_u$ and $D_d$ is more delicate, as demonstrated by \difcases{U0}, suggesting that a more forward-peaked DF in the downstream might saturate the limit (\ref{eq:s_limit}).

\begin{figure}
	\centerline{\epsfxsize=8.4cm \epsfbox{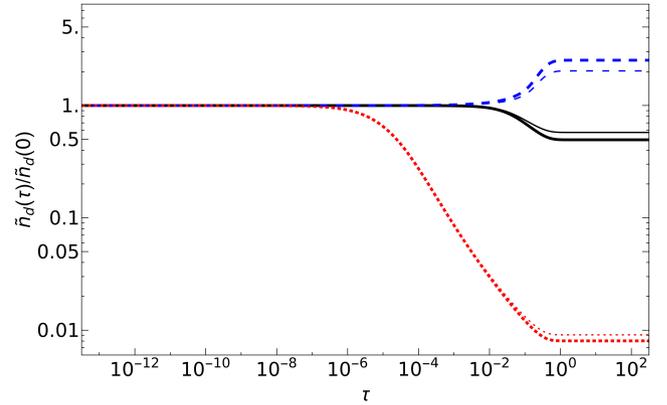}}
	\caption{Particle number density normalized to the shock front as a function of optical depth from the nominal shock. Shown are the cases \isoc{} (solid black), \difcases{U1} (dashed blue), and \difcases{D0} (dotted red), both in downstream frame (solid curves) and in the shock frame (where particle energy is not conserved; thin curves).
	}
	\label{fig:spatial_dis}
\end{figure}

Denote the particle density at optical depth $\tau$ from the shock by $\tilde{n}(\tau)\equiv\int_{-1}^1 \tilde{q}(\tilde{\mu},\tau)d\tilde{\mu}$, and consider the ratio between the downstream-frame densities far-downstream and at the shock front,
\begin{equation} \label{eq:xi}
\tilde{\xi} \MyEquiv_d \frac{\tilde{n}(\tau\rightarrow\infty)}{\tilde{n}(\tau=0)} \MyEqual_d 1+\frac{\int_{-1}^1 \tilde{\mu}\, \tilde{q}_s(\tilde{\mu})\,d\tilde{\mu}}{\beta \int_{-1}^1 \tilde{q}_s(\tilde{\mu})\,d\tilde{\mu}}\, ,
\end{equation}
where we used the conservation of $j$ in the last equality.
Figure \ref{fig:spatial_dis} shows the downstream evolution of the density ratio for a few choices of DFs.
For isotropic diffusion, $n(\tau)$ declines downstream of the shock, saturating at $\tilde{\xi}\simeq 0.5$ beyond $\tau\simeq 1$.
In contrast, for forward-beamed distributions, the last term in Eq.~(\ref{eq:xi}) is typically positive, so $\tilde{\xi}$ exceeds unity.
For example, the focused beam in case \difcases{U1} carries particles that isotropize only at some $\tau>0$, leading to a monotonically  increasing $n(\tau)$ that saturates at $\tilde{\xi}\simeq 2.6$.
We conclude that if a fraction $f_{acc}$ of the incoming energy is deposited in accelerated particles at the shock front, then more than double, $f_{acc}\tilde{\xi}$ of this energy is carried by the particles far downstream.
Similar results are obtained in the shock frame ($\xi\simeq 0.6$ for \isoc{} and $\xi\simeq1.9$ for \difcases{U1}; thin curves), but here particle energy is not conserved.

\subsection{Modified downstream diffusion raising $\myPret$}
\label{sec:hard_spec_pret}

Equations (\ref{pret}) and (\ref{eq:pesc}) suggest that $\myPret$ is sensitive to the downstream DF, which governs the fraction of particles returning to the shock after earlier crossing it downstream.
The spectrum hardens as more particles are deflected back toward the upstream, thus raising $\myPret$, which approaches unity for $j_-+j_+\ll j_+$, and typically also $\energain$.
Therefore, here we consider downstream DFs $D_d(\mu)$ that push particles backward, towards $\mu<0$ directions.

Again, suffice to demonstrate the behaviour by solving the problem for a selection of representative downstream DFs, keeping $D_u$ isotropic for the moment.
Taking a linearly decreasing (increasing) DF $D_d(\mu_d)$ now raises (lowers) both $\energain$ and $\myPret$, thus hardening (softening) the spectrum.
We confirm this behaviour using linear choices of $D_d$, finding an effect stronger than in the analogous test upstream, with the spectral index varying by $|\Delta s|\lesssim 0.3$ \citep{Keshet06}.

\begin{figure}
    \centerline{\epsfxsize=8.4cm \epsfbox{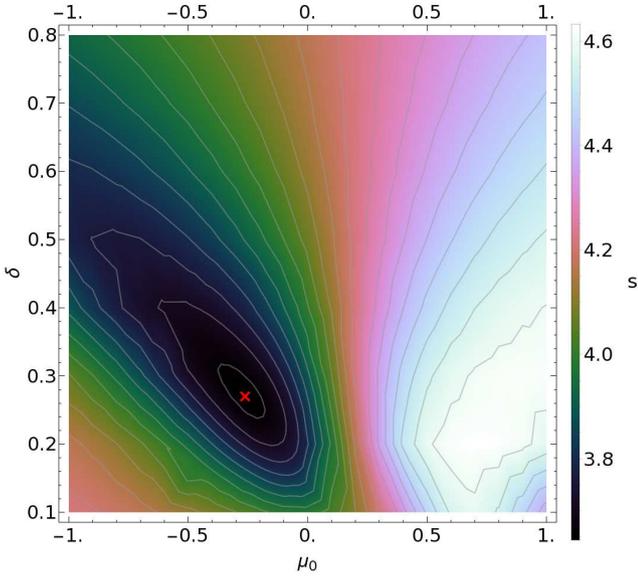}}
    \caption{Same as Fig.~\ref{fig:upstream_parameters} but for $D_d=D_G$ and isotropic diffusion upstream
    (here, contour intervals are $\Delta s=0.05$ starting at $s=3.65$, and $N=25$).
    The hardest case is marked (red cross; $s\simeq3.64$ with $\mu_0=-0.26$ and $\delta=0.27$;  case \difcases{D2} of Table~\ref{diff_table_down}).
    \label{fig:downstream_parameters}
    }
\end{figure}

Here too, a stronger effect is obtained by considering a localised change in the DF.
Figure \ref{fig:downstream_parameters} shows a scan of $D_d=D_G$ parameters with $h=100$ and an isotropic $D_u$, analogous to Fig.~\ref{fig:upstream_parameters}.
The spectrum generally hardens with decreasing $\mu_0$ and $\delta$, up to a valley along $2\mu_0+6\delta\simeq1$.
The hardest spectrum, $s\simeq3.64$, corresponds to $\mu_0=-0.25$ and $\delta=0.27$ (case \difcases{D2}).
A similar scan with $h=10^4$ shows a deeper valley, with the hardest spectrum, $s\simeq3.44$ at $\mu_0\simeq -0.25$ and $\delta\simeq 0.20$ (case \difcases{D1}).

When the spectrum is very hard due to a large $\myPret$, rather than a large $\energain$, the PDF does not become strongly peaked, and the problem can be accurately solved with fewer moments.
This allows us to more easily test DFs that are anisotropic both upstream and downstream.
For example, the same downstream diffusion of \difcases{D1} combined with $D_u=\exp(-b\mu)$, leads to even harder spectra: $s\simeq3.3$ for $b=1$, and $s\simeq3.05$ for $b=15$, labeled \difcases{U0}.

\begin{figure}
\centerline{\epsfxsize=8.4cm \epsfbox{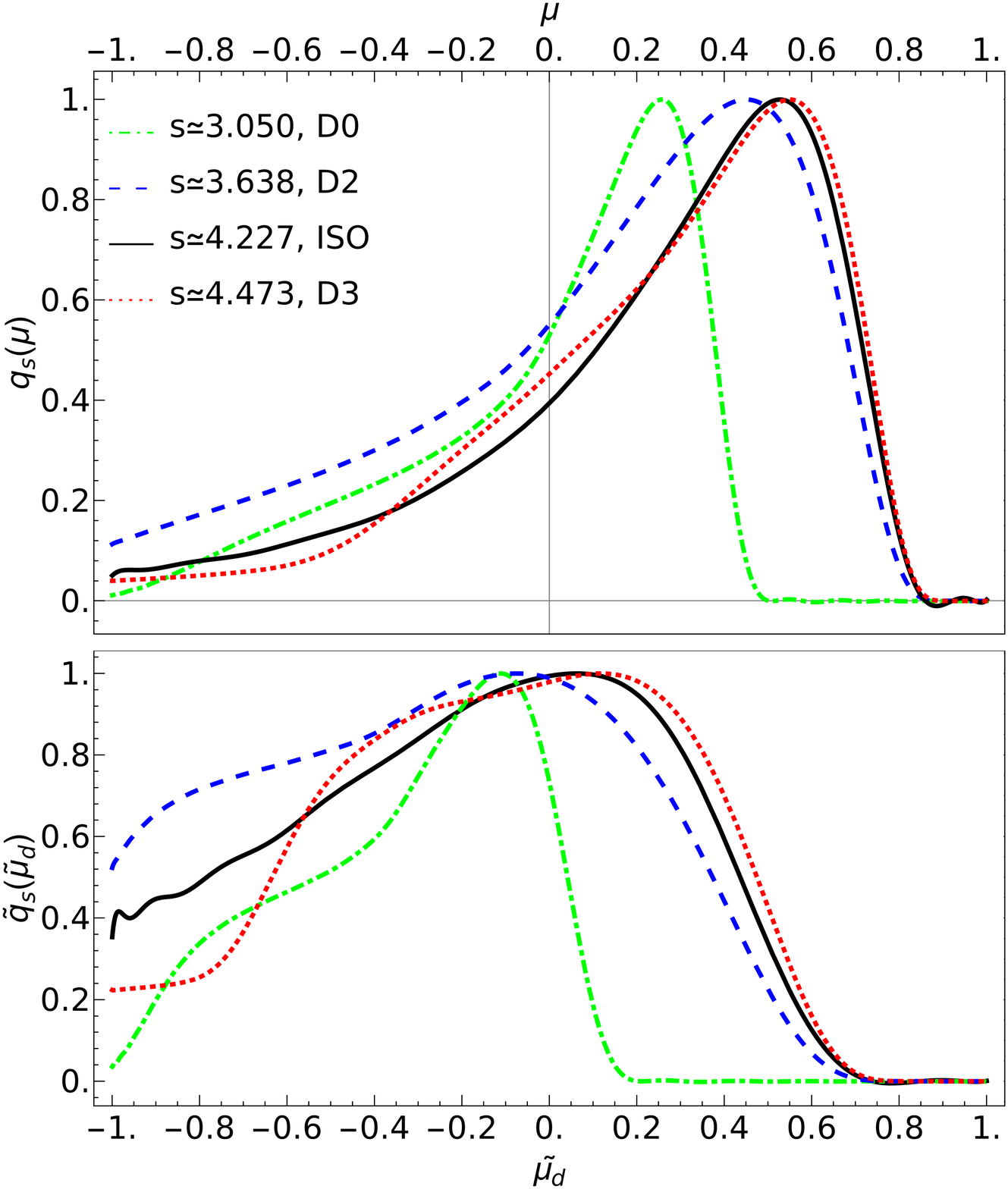}}
    \caption{Same as Fig.~\ref{fig:pdf_cases_up} in the shock (top) and downstream (bottom) frames, but with different return probabilities in downstream DF cases \isoc{} (solid black), \difcases{D0} (dot-dashed green), \difcases{D2} (dashed blue), and \difcases{D3} (dotted red). For the details of each case see Table~\ref{diff_table_down}.
    }
    \label{fig:pdf_cases_down}
\end{figure}

The angular distributions $q_s(\mu)$ for three representing cases of anisotropic downstream diffusion are shown in Fig.~\ref{fig:pdf_cases_down}.
The functional form of $q_s(\mu)$ does not vary much among these cases, showing a pronounced mode with a slightly varying $0.2\lesssim\mu_{\mbox{\tiny max}}\lesssim0.5$ position, anti-correlated with $s$ as expected from the dominant effect of $\myPret$ on the spectrum.
The downstream-frame distributions $\tilde{q}_s(\tilde{\mu})$ show the same anti-correlation between $\mu_{\mbox{\tiny max}}$ and $s$, combined with a narrowing of the particles angular distribution around that direction, as the spectrum hardens.

\begin{figure}
	\centerline{\epsfxsize=8.4cm \epsfbox{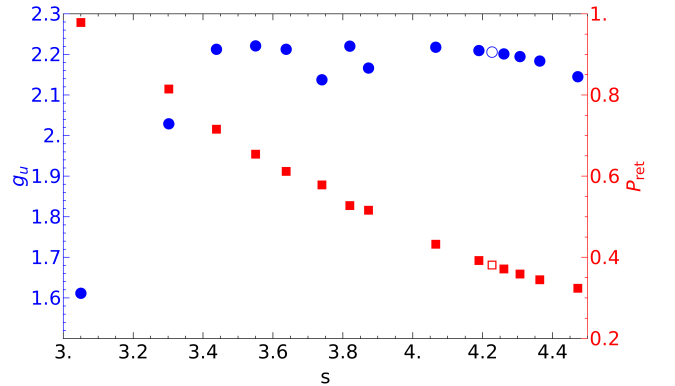}}
	\caption{Same as Fig.~\ref{fig:pret_s_up} but for the 14 examined downstream diffusion cases (see Table~\ref{diff_table_down}).}
	\label{fig:pret_s_down}
\end{figure}

Figure \ref{fig:pret_s_down} shows the values of $\energain$ and $\myPret$ for different choices of downstream DFs.
While $\energain\sim2.2$ does not vary much among most cases, $\myPret$ changes substantially, increasing for harder spectra, with the hardest case \difcases{D0} showing a near unity $\myPret\sim0.98$.
The hardest two spectra, obtained by considering anisotropic choices of $D_u$, manage to raise $\myPret$ at the cost of lowering $\energain$.

Finally, consider the downstream evolution of $\tilde{n}(\tau)$ for such cases, where the spectrum approaches its hard limit due to a near-unity $\myPret$.
Inspection of the bottom panel of  Fig.~(\ref{fig:pdf_cases_down}) shows that the PDF is almost entirely concentrated at $\tilde{\mu}_d<0$, so Eq.~(\ref{eq:xi}) allows for a very small $\tilde{\xi}$.
Indeed, Fig.~\ref{fig:spatial_dis} shows (dotted red curves) that for \difcases{D0}, $\tilde{n}(\tau)$ decreases substantially, saturating at $\tilde{\xi}\simeq 0.008$ for $\tau\gtrsim 1$.
Assuming that $f_{acc}$ does not exceed unity at the shock front, we conclude that the acceleration efficiency measured downstream is bounded here by $f_{acc}\tilde{\xi}\ll1$.
Characteristic numbers for $\tilde{\xi}$ as a function of $s$ are $0.17$, $0.25$ and $0.48$ for $s\simeq3.44$, $3.64$, and $4.19$, respectively (see Table~\ref{diff_table_down} for all other cases).
These results limit the applicability of high $\myPret$ models for particle acceleration that is both very hard and efficient.

\section{Summary and discussion}
\label{sec:summary}

We study the radio emission from PWNe in search of clues for the mechanism responsible for the unusually hard spectrum.
A phenomenological review of PWNe from the literature (\S\ref{sec:pwne}) suggests that the termination shock accelerates an $s\simeq 3.0$ electron spectrum. As this value is the hard spectral limit of Fermi processes such as DSA (\S\ref{sec:method}), we show that the latter produces sufficiently hard spectra under rather extreme conditions on the angular diffusion function of the accelerated particles (\S\ref{sec:hard_spectrum}), either upstream or downstream of the shock.

We compile a comprehensive sample of all 29 usable radio spectra from the literature (Table~\ref{pwn_table} and Fig.~\ref{fig:histogram}).
The sample is fairly well fit by a normal, $\alpha=0.20\pm0.20$ distribution of radio spectral indices.
Two soft, $\alpha\simeq 0.6$ outliers appear to be statistically distinguishable from the main sample; removing them leaves an $\alpha=0.16\pm0.16$, approximately normal, distribution (\S\ref{subsec:Model}).
More importantly, we find that in core-type PWNe such as CTB 80 (Fig.~\ref{fig:coreexample}), in which the spectrum varies spatially, the termination-shock region shows a harder, $\alpha=0.01\pm0.06$ distribution (Fig.~\ref{fig:corehistogram}), distinguishable from the remaining sample at a nominal $\sim3\sigma$ confidence level (\S\ref{sec:classifications}).

The harder radio spectra of the core-type group appear to be directly associated with the acceleration of electrons, whereas the highly uniform spectra of Crab-type PWNe could be affected by evolutionary effects as the cooling time (\ref{eq:sych_cool}) is long.
The sample is small and complicated by systematic effect; some evidence that core-type PWNe are older and more often associated with bow-shocks is only marginal (\S\ref{subsec:Implications}). More data will allow us to test the classification and its physical origin.

The $s\simeq 3$ electron spectrum inferred from PNWe, and in particular from the core-type sub-sample, plays an important role in the Fermi acceleration process and its manifestation in the DSA mechanism, as the hardest spectrum possible (Eq.~\ref{eq:s_limit}). Unless $\alpha=0$ is shown to be a natural outcome of some competing acceleration mechanism, our results thus suggest that (i) DSA might be responsible for the acceleration of the radio-emitting electrons; and (ii) DSA nearly saturates its hard spectral limit for the physical conditions around the PWN termination shock.

While linear DSA with large-angle particle scattering was previously shown to generate very hard spectra under some circumstances, it is unclear if it can approach the hard, $s=3$ limit without fine-tuning.
It was similarly unknown until now if small-angle scattering can so dramatically deviate from the $s\simeq 4.2$ spectrum typical of relativistic shocks.
We show that spectra arbitrarily close to the limit can be obtained for a sufficiently anisotropic angular DF either upstream or downstream, due to a corresponding increase in $\energain$ or $\myPret$.
These two diagnostics, sufficient for determining the spectrum (Eq.~\ref{eq:Fermi_relation}), can be accurately inferred even in the ultra-relativistic shock limit, from the PDF reconstructed by a moment expansion \citep{Keshet06}, provided that $\energain$ is measured upstream (Fig.~\ref{fig:energain_spectrum}).

Our results confirm previous suspicions \citep{Keshet06} that sufficiently anisotropic DFs can lead to substantial spectral deviations.
Thus, an upstream DF peaked in forward, $\mu\simeq 0.9$ angles (Fig.~\ref{fig:upstream_parameters}) can substantially raise $\energain\gg1$ (Fig.~\ref{fig:pret_s_up}), pushing the particles toward larger $\mu$ (Fig.~\ref{fig:pdf_cases_up}) and thus hardening the spectrum near its limit (Table \ref{diff_table_up}), with the particle density increasing downstream (Fig.~\ref{fig:spatial_dis}).
An analogous effect is obtained by changing the DF downstream.
Here, a downstream DF peaked in backward, $\mu\simeq -0.25$ angles (Fig.~\ref{fig:downstream_parameters}) can substantially raise $\myPret\simeq 1$ (Fig.~\ref{fig:pret_s_down}), pushing the particles toward smaller $\mu$ (Fig.~\ref{fig:pdf_cases_down}) and thus hardening the spectrum to its limit (Table \ref{diff_table_down}), with the particle density rapidly decreasing downstream (Fig.~\ref{fig:spatial_dis}).

Regardless of the acceleration mechanism, it is interesting to ask why do the termination shock regions of core-type PWNe show a harder spectrum than other types of nebulae.
One possibility is that the acceleration region has been evacuated from softer electrons, which were accelerated earlier, diffused inward, or otherwise evolved.
An alternative explanation is that some change in the environment of these termination shocks has modified the properties of the acceleration process.
In the context of DSA, such a modification could take place downstream and even upstream of the shock, if the accelerated electrons themselves play an important role in the dynamics.

If the radio electrons are accelerated by an extreme form of DSA, with highly anisotropic scattering, then the high $\energain$ variant seems more likely than the high $\myPret$ route, for two reasons.
First, in the former scenario, in which the highly anisotropic diffusion is attributed to the upstream, we find a strong anti-correlation between the PDF and the DF upstream, which could in principle arise non-linearly as the electrons generate the modes that scatter themselves \citep[\eg][]{nagar2019diffusive}.
Second, here the particle density grows downstream, so the acceleration efficiency is not bounded as it is for highly anisotropic downstream diffusion.

Nevertheless, we cannot rule out the alternative, large $\myPret$ option based on efficiency arguments.
The prototypical Crab nebula has an integrated synchrotron luminosity of about $1/4$ of the spin-down luminosity \citep{hester2008crab}, implying a high efficiency, but the spectrum here is not so hard; it is comparable to case \difcases{D2} with $s\simeq3.64$, where $\tilde{\xi}\sim25\%$.
A hard spectrum as found in CTB 80 is comparable to case \difcases{D0} with $\tilde{\xi}\sim0.8\%$, but here the efficiency may indeed be low, as inferred by comparing the radio luminosity $\lesssim10^{33}~\mbox{erg s}^{-1}$ and x-ray luminosity $\sim1.4\times10^{34}~\mbox{erg s}^{-1}$ \citep[][and references therein]{hester1988origin} with the spin-down luminosity $\sim3.7\times10^{36}~\mbox{erg s}^{-1}$ \citep{kulkarni1988fast}.

\section*{Acknowledgements}

We thank Y. Lyubarsky, Y. Nagar, and Y. Naor for helpful discussions.
This research has received funding from the GIF (Grant No. I-1362-303.7/2016), from an IAEC-UPBC Joint Research Foundation Grant (No. 300/18), and from the Israel Science Foundation (Grant No. 1769/15), and was supported by the Ministry of Science, Technology \& Space, Israel.







\bibliographystyle{mnras}
\bibliography{DSA}




\appendix


\section{Examined cases}
\label{sec:appendix_cases}

Our different choices of anisotropic DFs are summarised in Table \ref{diff_table_up} focusing on  variable  $\energain$, and in Table \ref{diff_table_down} focusing on  variable  $\myPret$, along with the resulting spectrum and DSA diagnostics, for the nominal shock.

\begin{table*}
    \caption{Examples of anisotropic DFs primarily upstream, raising $\energain$ for the nominal shock.
    }
    \centering
    \begin{tabular}{lllcccccccc}
        \hline
        Label&$D_u(\mu)$&$D_d(\mu)$&$s$&$s-s_{iso}$&$\mu_{\mbox{\tiny max}}$&$\myPret$&$\energain_u$&$s(\myPret,\energain_u)$&$\tilde{\xi}$&$N$\\
        (1) & (2) & (3) & (4) & (5)& (6)& (7)&(8)&(9)&(10)&(11)\\
        \hline
        \difcases{U0}&$D_G(\mu,10^6,0.9,0.2)$&$\exp(\mu)$&3.3034&-0.9235&0.999&0.169&354.2&3.3027&2.155&120\\
        \difcases{U1}&$D_G(\mu,10^6,0.9,0.2)$&1&3.3657&-0.8613&0.999&0.118&373.0&3.3611&2.532&120\\
        -&$D_G(\mu,10^6,0.9,0.3)$&1&3.4547&-0.7723&$\simeq1$&0.125&136.2&3.4234&2.389&95\\
        &$D_G(\mu,10^3,0.9,0.3)$&1&3.7221&-0.5050&0.966&0.150&15.01&3.7012&2.055&40\\
        \difcases{U2}&$D_G(\mu,10^2,0.9,0.2)$&1&3.8673&-0.3597&0.935&0.189&7.080&3.8500&1.478&40\\
        -&$D_G(\mu,10^2,0.9,0.4)$&1&3.9448&-0.2823&0.898&0.208&5.344&3.9356&1.390&25\\
        -&$D_G(\mu,10^2,0.9,0.5)$&1&4.0191&-0.2079&0.857&0.241&4.057&4.0159&1.124&20\\
        -&$1+0.9\mu$&1&4.1805&-0.0465&0.647&0.339&2.494&4.1823&0.622&20\\
        \isoc{}&1&1&4.2270&0&0.527&0.379&2.208&4.2251&0.495&15\\
        \difcases{U3}&$1-0.9\mu$&1&4.2640&0.0370&0.447&0.416&2.011&4.2565&0.402&20\\
        \hline
    \end{tabular}
    \label{diff_table_up}
    \vspace{0.2cm}
    \begin{tablenotes}
        \item Columns: (1) Case label; (2) Upstream DF; (3) Downstream DF; (4) Electron spectral index calculated using the moments method; (5) Spectral difference with respect to the case of isotropic DF; (6) Angular location of PDF maximum; (7) Return probability; (8) Energy gain calculated in the upstream frame; (9) Spectral index calculated based on Eq.~(\ref{eq:Fermi_relation}); (10) Ratio between the downstream-frame densities far-downstream and at the shock front; (11) The number of moments used in the computation.
    \end{tablenotes}
\end{table*}

\begin{table*}
	\centering
	\caption{
	Same as Table~\ref{diff_table_up} but for anisotropic DFs predominantly downstream, raising $\myPret$.}
	\begin{tabular}{lllcccccccc}
		 \hline
        Label&$D_u(\mu)$&$D_d(\mu)$&$s$&$s-s_{iso}$&$\mu_{\mbox{\tiny max}}$&$\myPret$&$\energain_u$&$s(\myPret,\energain_u)$&$\tilde{\xi}$&$N$\\
        (1) & (2) & (3) & (4) & (5)& (6)& (7)&(8)&(9)&(10)&(11)\\
        \hline
		\difcases{D0}&$\exp(-15\mu)$&$D_G(\mu,10^{4},-0.25,0.2)$&3.0504&-1.1766&0.257&0.977&1.614&3.0494&0.008&45\\
		-&$\exp(-\mu)$&$D_G(\mu,10^4,-0.25,0.2)$&3.3022&-0.9250&0.336&0.813&2.031&3.2926&0.096&35\\
		\difcases{D1}&1&$D_G(\mu,10^4,-0.25,0.2)$&3.4384&-0.7886&0.419&0.714&2.215&3.4242&0.171&40\\
		-&1&$D_G(\mu,10^4,-0.3,0.3)$&3.5504&-0.6766&0.441&0.652&2.223&3.5349&0.219&25\\
		\difcases{D2}&1&$D_G(\mu,10^2,-0.26,0.27)$&3.6379&-0.5891&0.450&0.610&2.215&3.6220&0.252&25\\
		-&1&$D_G(\mu,10^4,0,0.2)$&3.7402&-0.4868&0.429&0.576&2.140&3.7242&0.263&45\\
		-&1&$D_G(\mu,10^4,-0.4,0.5)$&3.8200&-0.4069&0.480&0.526&2.223&3.8055&0.331&20\\
		-&1&$D_G(\mu,10^4,0,0.3)$&3.8736&-0.3534&0.469&0.514&2.169&3.8595&0.327&25\\
		-&1&$1-0.7\mu$&4.0521&-0.1610&0.512&0.430&2.220&4.0574&0.435&15\\
		-&1&$1-0.2\mu$&4.1894&-0.0376&0.525&0.390&2.212&4.1857&0.483&15\\
        \isoc{}&1&1&4.2270&0&0.527&0.379&2.208&4.2251&0.495&15\\		-&1&$1+0.2\mu$&4.2609&0.0339&0.530&0.369&2.204&4.2606&0.507&15\\
		-&1&$1+0.5\mu$&4.3070&0.0800&0.537&0.357&2.197&4.3092&0.522&20\\
		-&1&$1+0.9\mu$&4.3636&0.1366&0.542&0.343&2.186&4.3692&0.537&20\\
		\difcases{D3}&1&$D_G(\mu,10^4,0.3,0.2)$&4.4725&0.2455&0.550&0.322&2.147&4.4836&0.541&40\\
		\hline
	\end{tabular}
	\label{diff_table_down}
    \vspace{0.2cm}
\end{table*}


\label{lastpage}
\end{document}